\documentclass[pra, aps, groupedaddress, superscriptaddress, notitlepage, reprint, 10pt]{revtex4-2}

\usepackage[utf8x]{inputenc}
\usepackage{color}
\usepackage{bbm} 

\usepackage{nicefrac}
\usepackage{amsfonts,amsmath,amssymb,stmaryrd}
\usepackage{braket}
\usepackage{mathtools}
\usepackage[autostyle=true]{csquotes}
\usepackage{accents}

\usepackage{colonequals}
\usepackage{multirow} 
\usepackage{hhline}
\usepackage{graphicx}
\usepackage{subfig}  
\usepackage{bbm} 
\usepackage[english]{babel}
\usepackage[hidelinks]{hyperref}
\addto\extrasenglish{
}

\usepackage[pdftex]{epsfig}
\usepackage{mathrsfs}
\usepackage{verbatim}
\usepackage{centernot}
\usepackage[normalem]{ulem}
\usepackage{cancel}
\usepackage{ifthen}
\usepackage{bm}	
\usepackage{siunitx}
\usepackage{float} 
	\usepackage{dsfont}
	\usepackage[super]{nth} 
	\usepackage{todonotes}
	\InputIfFileExists{gitinfo-latex.inc}{}{}
	
	\usepackage{listings}
	\usepackage{color}

\newcommand{\Cs}{\textsuperscript{133}Cs \,}
\newcommand{\Rb}{\textsuperscript{87}Rb \,}

\DeclareSIUnit\gauss{G}	
\DeclareSIUnit\Kelvin{K}

\usepackage{ragged2e}
\makeatletter
\renewcommand{\@makecaption}[2]{%
	\vskip\abovecaptionskip
	\begin{minipage}{\linewidth}
		\justifying
		#1. #2
	\end{minipage}
	\vskip\belowcaptionskip
}
\makeatother

\newif\ifmeasindexsup
\measindexsupfalse  

\newcommand{\measidx}[1]{%
	\ifmeasindexsup
	^{(#1)}%
	\else
	_{#1}%
	\fi
}

\begin{document}
	
	\title{Bayesian Quantum Thermometry of Ultracold Gases with \\ Non-Equilibrium Single-Atom Probes}
	\author{Julian Feß}
\author{Luca Granert}
\author{Sabrina Burgardt}
\author{Silvia Hiebel}
\author{Alexander Guthmann}
\affiliation{Department of Physics and State Research Center OPTIMAS, RPTU University Kaiserslautern-Landau, 67663 Kaiserslautern, Germany}
\author{Mohammad Mehboudi}
\affiliation{Vienna Center for Quantum Science and Technology, Atominstitut, TU Wien, Vienna 1020, Austria}
\author{Martí Perarnau-Llobet}
\affiliation{Física Teòrica: Informació i Fenòmens Quàntics, Departament de Física, Universitat Autònoma de Barcelona, 08193 Bellaterra (Barcelona), Spain}
\author{Artur Widera}
\affiliation{Department of Physics and State Research Center OPTIMAS, RPTU University Kaiserslautern-Landau, 67663 Kaiserslautern, Germany}

	\date{20 November 2026}
	\begin{abstract}
		Precise thermometry is essential for experiments with ultracold quantum gases, but conventional techniques such as time-of-flight thermometry are destructive and lose sensitivity at low atomic densities. Quantum probes provide an alternative route by encoding temperature information in accessible internal degrees of freedom. Here, we implement Bayesian thermometry with non-equilibrium spin-based single-atom probes immersed in an ultracold atomic bath. Temperature-dependent spin-exchange collisions map the bath temperature onto the internal spin populations of the probe atoms. Measurements of these populations are then used to iteratively update a Bayesian temperature estimate. We connect the information gained at different measurement settings to the underlying spin dynamics, identifying intersections of temperature-dependent population trajectories as a source of thermometric ambiguity. Using the expected information gain to optimize measurement settings, we experimentally compare unoptimized, a~priori optimized, and adaptively optimized protocols. Both optimized protocols accelerate convergence, with the adaptive protocol reducing the average number of measurements required to reach a prescribed precision by a factor of 2.5 compared with the unoptimized protocol. In the many-measurement regime, we further characterize the protocols through their asymptotic error scaling and the accumulation of Fisher information, while testing consistency with the expected Cramér-Rao bound (ECRB) and the Bernstein-von Mises theorem. These results connect the microscopic dynamics of a quantum probe to efficient measurement design and establish Bayesian spin thermometry as a practical approach for ultracold-gas experiments.
		
	\end{abstract}
	\maketitle
	%
	
	\section{Introduction}
\label{sec:introduction}

Temperature is a key parameter for characterizing and controlling ultracold quantum gases.
In most setups, the standard approach for determining the temperature is time-of-flight thermometry, where the gas is released from the trap and its expansion is imaged after a period of free evolution. While highly successful, this method is destructive and becomes increasingly challenging for small samples, low atomic densities, or scenarios requiring repeated measurements on the same system. These limitations motivate non-invasive measurement schemes based on quantum probes \cite{Mehboudi2019Thermometry}.

Quantum probes are highly controllable quantum systems that can be coupled to a sample to infer its temperature and other physical properties. By preparing and monitoring the probe's state, information about the sample can be extracted either at thermal equilibrium or during transient non-equilibrium dynamics \cite{Mehboudi2019Thermometry}. Experimentally, single-atom probes have been realized to estimate the temperature \cite{SpinThermometry}, magnetic fields \cite{Nettersheim2023Sensitivity, Schaffner2024Jan}, and bath density \cite{Adam2022Coherent, Schmid2010Sep} of ultracold gases, while single-qubit probes have also been implemented in superconducting circuits \cite{Lvov2025Thermometry}. In parallel, theoretical efforts have established the fundamental precision limits of quantum probes both in equilibrium \cite{Correa2015Individual,Potts2019,Jrgensen2020Tight,Abiuso2024Optimal,Ullah2026Configuration} and non-equilibrium regimes \cite{Hovhannisyan2021,Mehboudi2025Optimal,Sekatski2022Optimal}, together with novel thermometry schemes based on coherent dynamics \cite{Mitchison2020,Mitchison2022,Albarelli2023Invasiveness,Ullah2023,Boeyens2023Probe,Brattegard2024,Aiache2024,AntoSztrikacs2024,Trombetti2026Nonequilibrium,Boudjema2024Quantum,Ullah2025Single} and strong probe-sample coupling \cite{Correa2017,Hovhannisyan2018,Mehboudi2019Using,Zhang2021Non,Planella2022Bath,Khan2022Subnanokelvin,Luiz2022Machine,Yuan2023Quantum,RavellRodrguez2024,Aiache2024,winczewski2026non}.

To date, most advances in quantum thermometry have focused on the frequentist approach to address local estimation and the limit of large data. However, when experimental data are scarce, or in global estimation, a Bayesian approach is better suited \cite{Mukhopadhyay2025}. This has motivated the development of Bayesian thermometry protocols and precision limits \cite{GlobalThermometry,Boeyens2021Uninformed,Jrgensen2022Bayesian,Mehboudi2022Fundamental,Rubio2022Quantum,Bavaresco2024,Nimmrichter_Rubio_symmetry_geometry,Andr2026Strategy}. Bayesian estimation has recently begun to move from theory into cold-atom experiments, where it has been used to speed up release–recapture thermometry of few trapped atoms in tweezers \cite{Glatthard2022Optimal} and to improve atom-number measurements \cite{Overton2026Adaptive}. Here, we bring Bayesian thermometry to a probe with rich internal dynamics: single-atom spin probes immersed in an ultracold bath~\cite{SpinThermometry,Nettersheim2023Sensitivity}, where temperature is encoded in multilevel, non-equilibrium spin dynamics.

Our system uses individually measured \Cs atoms with quantized spin states as probes, which are coupled to a thermal \Rb cloud via elastic and inelastic collisions \cite{Tailored_Single-Atom_Collisions}. Temperature is inferred from the non-equilibrium spin dynamics of the \Cs atoms induced by inelastic spin-exchange collisions \cite{SpinThermometry, Nettersheim2023Sensitivity}. Since the optimal choice of measurement parameters, namely the interaction time and the measured spin state, depends on the unknown temperature, the problem can benefit highly from adaptive protocols, which the Bayesian approach enables straightforwardly. Using the Bayesian mean information gain per measurement as an optimization criterion, we design and compare three distinct measurement protocols that differ in their degree of optimization: an unoptimized protocol, an a~priori optimized protocol, and a greedy adaptive protocol. Of these schemes, the greedy adaptive protocol makes the most direct use of the Bayesian framework, as it incorporates a feedback loop, as sketched in \autoref{fig:conceptual_figure}(a). While the adaptive scheme achieves the highest performance, reducing the number of measurements required to reach a pre-set precision by a factor of 2.5, our results highlight a key trade-off between measurement sensitivity and experimental overhead.
We also investigate the asymptotic limit of many measurements. Remarkably, our estimation protocols, in all three scenarios, closely follow the expected Cram\'er-Rao bound (ECRB). Moreover, the posterior distribution is consistent with the Bernstein-von Mises theorem, which states that its variance is inversely proportional to the accumulated Fisher information.
Ultimately, this work establishes Bayesian single-atom spin thermometry as a versatile, high-precision tool for ultracold gas experiments, especially when data acquisition is constrained or costly.

	\section{Methods}
\label{sec:methods}

We first establish the working principle of our quantum probe. We begin with a microscopic description of the physical system, consisting of ultracold bath atoms and atomic impurities, and discuss their interaction. We then introduce the non-equilibrium spin dynamics induced by inelastic spin-exchange collisions and show how these dynamics determine the spin-state occupation probabilities. These probabilities are connected to the experimentally measured atom numbers through the atom-number likelihood used for Bayesian inference. Next, we describe the Bayesian estimation scheme and the information gain associated with different measurement parameters. Finally, we combine the spin dynamics with Bayesian inference to realize three measurement protocols that differ in their degree of optimization.

\subsection{Non-Equilibrium Spin Dynamics}
\label{subsec:SE_dynamics}
\begin{figure*}
	\includegraphics[width=\linewidth]{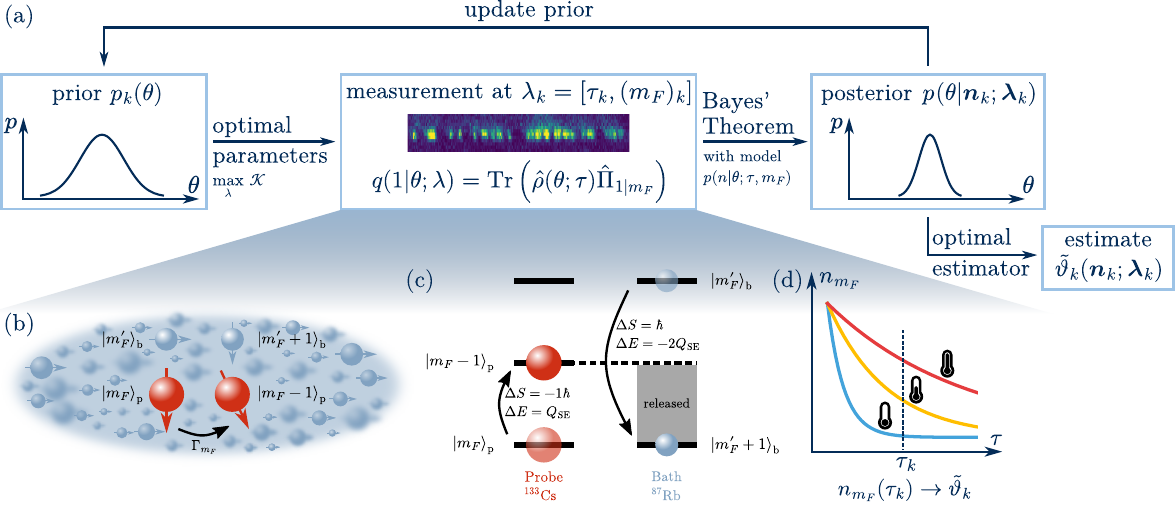}
	\caption{\textbf{Realization of the probe--bath system and Bayesian inference.} (a) Illustration of the Bayesian feedback loop. A prior probability density function $p_{k}(\theta)$ encodes initial knowledge and is used to calculate optimal measurement parameters by maximizing the mean information gain $\mathcal{K}$. Depending on the protocol, a measurement is performed at the (optimal) parameters $\lambda_k$. The image shows a fluorescence image of single probe atoms as an example of a single measurement. The measurement is a projective measurement that detects whether a probe atom is in state $m_F$, corresponding to the POVM element $\hat\Pi_{1|m_F}=\ket{m_F}\bra{m_F}$, the projector onto the basis state $\ket{m_F}$. The occupation probability is therefore given by $q(1|\theta;\lambda)=\mathrm{Tr}(\hat\rho(\theta;\tau)\hat\Pi_{1|m_F})$. Using Bayes' theorem (\autoref{eq:bayes}) and a microscopic model of the probe-bath dynamics $p(n|\theta;\tau,m_F)$,  we obtain a posterior PDF $p(\theta|\boldsymbol{n}_k;\boldsymbol{\lambda}_k)$, which is typically narrower than the prior. This posterior is then used to calculate an estimate $\tilde{\vartheta}_k(\boldsymbol{n}_k;\boldsymbol{\lambda}_k)$ using a point estimator. The posterior becomes the prior of the next measurement. (b) Sketch of the atoms forming the probe--bath system. Multiple Cs atoms (red) interact with the thermal ultracold Rb bath (blue) via spin-exchange collisions at temperature-dependent rates $\Gamma_{m_F}(T)$. (c) Internal spin states of a probe atom (left) and a bath atom (right). The Zeeman splittings differ by a factor of two, given by the corresponding Landé factors. A spin-exchange collision transfers $1\hbar$ from the probe to the bath atom. The energy $2Q_\mathrm{SE}$ is released by the bath atom, of which the probe atom absorbs $Q_\mathrm{SE}$, while the remaining energy $Q_\mathrm{SE}$ is released into the kinetic degrees of freedom. This excess kinetic energy is rapidly dissipated by more frequent thermalizing elastic collisions. (d) Temperature sensitivity is rooted in different evolution speeds, and is thus mapped into different atom numbers at a fixed interaction time $\tau$.}
	\label{fig:conceptual_figure}
\end{figure*}
We utilize inelastic spin-exchange (SE) collisions to map the thermal information of an ultracold atomic bath onto the internal spin state of our single-atom quantum probe.
Experimentally, we realize the probe--bath system by immersing approximately 40 non-interacting and uncorrelated \Cs atoms into a thermal cloud of \Rb atoms, as sketched in \autoref{fig:conceptual_figure}(b), see also Sec. \ref{sec:experimental_sequence}. The bath consists of about $N_\text{b}\approx\num{6.6(0.3)e3}$ atoms, at temperatures $T$ between $\SI{500}{\nano\kelvin}$ and $\SI{1500}{\nano\kelvin}$. This corresponds to mean bath densities of $\langle n_\mathrm{b}\rangle_\mathbf{r}=(\num{0.54}\text{--}\num{1.58})\times\SI{e12}{\per\cubic\centi\meter}$. Each \Cs atom acts as an independent single-atom probe. Both atomic species, \Cs and \Rb, are laser cooled into their hyperfine ground states, $F_\text{p}=3$ and $F_\text{b}=1$, yielding $2F_\text{p,b}+1$ equally spaced Zeeman substates, i.e., seven and three states, respectively. These substates define the internal spin degrees of freedom. We denote these states by their magnetic quantum number $\ket{m_F}_\text{p,b}$. For the probe atoms, $^{133}\text{Cs}$, $m_F$ takes values from $+3$ to $-3$, from the absolute ground state to the highest-energy Zeeman state within $F_\text{p}=3$. These states are coupled via inelastic $s$-wave spin-exchange collisions between probe and bath atoms, which induce unidirectional nearest-neighbor transitions,
\begin{equation}
	\begin{aligned}
		\ket{m_F}_\text{p}\otimes\ket{m_F'}_\text{b}\xrightarrow[\text{coll.}]{\text{SE}}\ket{m_F-1}_\text{p}\otimes\ket{m_F'+1}_\text{b} + Q_\mathrm{SE},
	\end{aligned}
	\label{eq:SE_reaction}
\end{equation}
releasing the internal energy $Q_\mathrm{SE}$, as indicated in \autoref{fig:conceptual_figure}(c) \cite{Tailored_Single-Atom_Collisions}. The corresponding scattering cross sections depend on the collision energy and are therefore temperature dependent.

Each spin-exchange collision transfers $1\hbar$ of angular momentum from the probe to the bath and an energy $Q_\mathrm{SE}=\mu_\text{B}B/4$ from the bath to the probe, where $B$ is the external homogeneous magnetic field, see \autoref{fig:conceptual_figure}(c). Due to the mismatch of the Zeeman energy splittings, set by the atomic Landé factors, an additional energy $Q_\mathrm{SE}$ is released into the kinetic degrees of freedom of the scattered atoms. This is a consequence of energy and angular-momentum conservation, given that the Zeeman splitting of the bath atoms is twice that of the probe atoms. A transfer of $2\hbar$ and $2Q_\mathrm{SE}$ is forbidden because the bath atoms can absorb only one quantum of spin, as they are prepared in $\ket{m_F=0}_\text{b}$ and $m_F$ takes values only from $+1$ to $-1$. The reverse process is strongly suppressed, as the collision energy would have to exceed $Q_\mathrm{SE}$. Under our experimental conditions, this requirement is only met by the high-energy tail of the collision-energy distribution, as we set the external magnetic field to $B=\SI{460}{\milli\gauss}$, resulting in an energy level splitting that is about five times larger than the mean collision energy at the mean temperature of the probing interval. The suppression of the reverse process therefore breaks detailed balance, the principle that every process of energy transfer must be accompanied by an allowed reverse process \cite{Seifert2012StochasticThermodynamics}.

The change in bath spin and internal energy can be neglected for two reasons. First, the number imbalance between probe and bath atoms is large, by a factor of approximately 200, and second, the probe atom traverses the whole bath between two successive collisions. The mean free path of the probe atoms, $\Lambda=\SIrange{22}{85}{\micro\meter}$, is at least comparable to the axial extent of the bath, $\sigma_z=\SIrange{24}{42}{\micro\meter}$, and much larger than the radial extents, $\sigma_{x,y}=\SIrange{1.6}{2.9}{\micro\meter}$. The lower values of the mean free path and bath sizes correspond to the coldest bath temperatures, while the larger values correspond to the hottest bath temperatures. Thus, a bath atom that has participated in a spin-exchange collision is unlikely to do so again. This justifies the assumption that bath atoms remain in their initially prepared internal state, $\ket{m_F=0}_\text{b}$, and supports the Markov approximation, in which the bath is treated as memoryless. We therefore consider only the internal state of the probe and omit the corresponding index. The argument for neglecting the decrease in bath internal energy is analogous. The large mean free path compared to the cloud size also implies that the probe does not measure only a local temperature, but rather samples the bath temperature over the whole cloud.

The situation for the increased kinetic energy is different. Here, Markovianity holds because the elastic scattering rates are $\SIrange{19}{44}{}$ times larger than the inelastic ones, depending on the probe internal state. This leads to rapid dissipation of the excess energy into the bath, justifying the assumption of a thermalized probe and bath.

The resulting population dynamics induced by spin-exchange collisions pump the probe atoms into the state $\ket{-3}$. Although this populates the highest-energy probe state, the total internal energy of the combined probe-bath system decreases, because the bath atom lowers its energy by $2Q_\mathrm{SE}$, whereas the probe atom increases its energy only by $Q_\mathrm{SE}$ per collision. The initial internal state of the probe, which is the absolute ground state $\ket{+3}$ throughout this manuscript, is depleted, while the intermediate states $\ket{2} \text{ to } \ket{-2}$ are transiently populated, reaching maximum fractional populations of $\num{0.2}$ to $\num{0.3}$. 

\begin{figure}[t]
	\subfloat{\includegraphics[]{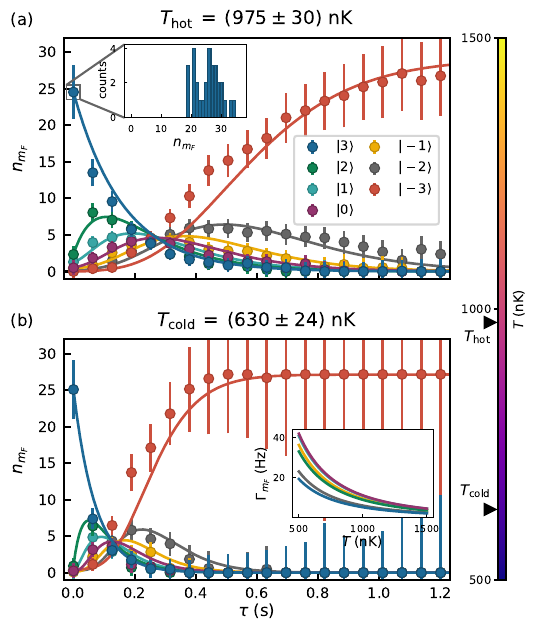}}\\
	\caption{\textbf{Spin-exchange collision induced population dynamics.} Absolute atom number $n_{m_F}$ of Cs atoms in state $\ket{m_F}$ (colors) after an interaction time $\tau$ (abscissa) with (a) a hot and (b) a cold Rb bath ($T_\text{hot}=\SI{975\pm30}{\nano\kelvin}$ and $T_\text{cold}=\SI{630\pm24}{\nano\kelvin}$). Solid lines show parameter-free simulations based on rate equations with rates obtained from coupled-channel scattering calculations. The inset of (b) displays the $\ket{m_F}$-state (color) and temperature dependence (abscissa) of those rates. The inset of (a) shows a histogram of the quantized atom-number detection for multiple measurements performed at identical parameters. Averaging over many such measurements yields the mean atom number represented by each data point. Error bars show Poissonian uncertainties.}
	\label{fig:SE_comparison}
\end{figure}

Exemplary state- and time-resolved measurements of SE evolutions for two different bath temperatures $T$ are shown in \autoref{fig:SE_comparison}, together with the scattering rates used to model these dynamics in the inset of panel (b). Each data point is the average of 15 independent fluorescence imaging measurements at identical parameter settings, as indicated by the histogram inset of \autoref{fig:SE_comparison}(a). To account for three-body losses, which lead to an exponential decay of the total probe atom number over interaction time, the data is rescaled such that the total atom number at each interaction time, $n_\text{tot}(\tau)\coloneqq\sum_{m_F}n_{m_F}(\tau)$, is fixed to its value at $\tau=0$, which is typically around 40. Here $n_{m_F}$ is the absolute atom number in state $\ket{m_F}$. In contrast to the rate-equation model, which yields occupation probabilities, the experiment measures integer atom numbers $n$. We therefore scale the modeled occupation probability by the initial total probe atom number $n_\text{tot}(0)$. Both datasets clearly exhibit the previously described pumping into $\ket{-3}$, differing primarily in the timescale of this process. The dynamics is faster for lower temperatures, as evidenced by the full polarization of the probe ensemble in $\ket{-3}$ after approximately $\SI{500}{\milli\second}$ for the cold bath, compared to about $\SI{1200}{\milli\second}$ for the hot bath.

This scaling of the evolution speed with the bath temperature arises from the dependence of the spin-exchange collision rates on the scattering cross sections $\sigma$ and the mean bath density $\langle n_\text{b}\rangle_\mathbf{r}$, with  $\mathinner{\langle \,\cdot\,\rangle}_\mathbf{r}$ denoting spatial average. The former depend on temperature implicitly through thermal averaging over the Maxwell-Boltzmann distribution, whereas the latter depends on it explicitly as $\langle n_\mathrm{b}\rangle\propto N_\text{b}\,T^{-3/2}$. Accordingly, the rates read
\begin{equation}
	\Gamma_{m_F}(T)=\mathinner{\langle\sigma(v_\text{c},m_F)\,v_\text{c}\rangle}_{T}\mathinner{\langle n_\text{b}(T, N_\text{b})\rangle}_\mathbf{r},
	\label{eq:SE_rates}
\end{equation}
with the collision velocity $v_\text{c}$, and $\mathinner{\langle\,\cdot\,\rangle}_T$ denoting the thermal average over the Maxwell--Boltzmann distribution of relative collision velocities (see Supplemental Material Sec.~\ref{subsec:MB}). Since both contributions decrease with increasing temperature, the resulting collision rates $\Gamma_{m_F}$ exhibit an overall decrease with temperature, as shown in the inset of \autoref{fig:SE_comparison}(b). Modeling these dynamics with rate equations requires the bath temperature $T$, bath atom number $N_\text{b}$, and the scattering cross sections $\sigma$. The former two are independently measured via absorption imaging of the bath after time-of-flight (TOF), while the latter are obtained from quantum-mechanical coupled-channel scattering calculations that account for the internal atomic structure of both species and the molecular interaction potential. 

To model the resulting non-equilibrium population dynamics, let $\bm q=\left(q_3,\,q_2,\,\ldots,\,q_{-3}\right)^\mathrm{T}$ denote the occupation-probability vector of the probe ensemble, where its elements $q_{m_F}$ represent the fractional population in state $\vert m_F\rangle$. Coherences between the Zeeman states are rapidly dissipated through interactions with the bath, such that the probe dynamics can be described in terms of the populations alone. The time evolution of the populations is described by the rate equation
\begin{align}\label{eq:rate}
	\dot{\bm q} = \bm \Gamma(T)\bm q.
\end{align}
Here, $\bm{\Gamma}$ is the rate matrix, whose temperature-dependent elements can be fully characterized using \autoref{eq:SE_rates}, namely
\begin{equation}
	\bm{\Gamma}_{m_F',m_F} =
	\begin{cases}
		-\Gamma_{m_F},
		& m_F'=m_F,\quad m_F\neq-3,\\
		\Gamma_{m_F},
		& m_F'=m_F-1,\quad m_F\neq-3,\\
		0, & \text{otherwise}.
	\end{cases}
\end{equation}
The column index denotes the initial state and the row index the final state of a spin-exchange collision. Solving the rate equations numerically yields modeled population probabilities without free fit parameters. The exact initial probe-state population vector $\bm q(\tau=0)$ serves as the initial condition. After scaling to absolute atom numbers, the modeled probabilities accurately reproduce the observed dynamics, as shown by the solid lines in \autoref{fig:SE_comparison}.

While the rate equations describe a continuous evolution of ensemble-averaged population probabilities, the underlying microscopic dynamics consist of discrete spin-exchange events that change the internal state of individual probe atoms. State-resolved fluorescence imaging then samples these discrete occupations and yields integer atom numbers. To use this description in our Bayesian inference framework, we construct a statistical likelihood model that translates the ensemble-averaged population probabilities obtained from the rate equations to the probabilities for observing integer atom numbers in a projective measurement, as we show in the next section.

\subsection{Atom detection, Bayesian inference, and information gain}
\label{subsec:global_quantum_thermometry}
We begin by clarifying our notation. Throughout this work, lowercase $p$ or $q$ denote probability density functions (PDF) or probability distributions, depending on the argument being discrete or continuous. A vertical bar ($|$) indicates conditional probability, and a semicolon ($;$) separates random variables from deterministic control parameters.
For example, we use $p(x, y | z, w; \lambda)$ to represent the probability of observing $x$ and $y$, conditioned on $z$ and $w$, for fixed control parameters $\lambda$. It should hold that $\sum_{x,y}p(x,y|z,w;\lambda) = 1$, whereas summing over only one of the variables gives the marginal conditional distribution of the other variable. Likewise, $p(z, w; \lambda)$ refers to the joint probability of $z$ and $w$ for a fixed $\lambda$. If the variables $z,w$ are independent of $\lambda$ one may represent this with $p(z,w)$. We have $\sum_{z,w}p(z, w; \lambda) = 1$, while summing over only one random variable gives the marginal distribution for the other one. 
Furthermore, while $\theta$ represents the hypothetical bath temperature, $T$ refers to the true but unknown bath temperature.
\subsubsection{Modeling the atom detection protocol}\label{subsubsec:model}
A single measurement of the probe atoms via our state-resolved fluorescence imaging detects whether a probe atom occupies a selected Zeeman state \(\ket{m_F}\). Figure \ref{fig:conceptual_figure}(a) shows such an exemplary fluorescence image. The probe state is diagonal in the Zeeman basis, thus the spin-exchange dynamics can be described by rate equations. Measuring a single population therefore corresponds to a coarse-grained projective measurement~\cite{NielsenChuang2010} with binary outcomes represented by the positive operator-valued measure (POVM) elements
\begin{align}
	\hat\Pi_{1|m_F} & =\ket{m_F}\bra{m_F},\nonumber\\
	\hat\Pi_{0| m_F} & =\mathds{1}-\ket{m_F}\bra{m_F}=\sum_{m_F'\neq m_F}\ket{m_F'}\bra{m_F'}.
	\label{eq:POVM}
\end{align}
In our notation above, $\hat{\Pi}_{a|x}\geq 0 ~\forall~\{a,x\},$ refers to the outcome $a$ of the $x$ measurement, and it holds that $\sum_a {\hat\Pi_{a|x}} = \mathds{1}~\forall~ x$. The probability that a \textit{single atom}  occupies the state $\ket{m_F}$ after interacting for time $\tau$ with a bath at temperature $\theta$ is then obtained via the Born rule
\begin{equation}
	q(1|\theta;\lambda)=\mathrm{Tr}\!\left(\hat\rho(\theta;\tau)\hat\Pi_{1|m_F}\right),
	\label{eq:occ_prob_from_POV}
\end{equation}
where we define $\lambda = (\tau, m_F)$ as the control parameters. Note that we have $q(1|\theta;\tau,m_F) \equiv q_{m_F}$ and thus the dynamical behavior of single-atom occupations is readily known thanks to \autoref{eq:rate}. Furthermore, while we use $q$ to refer to the single-atom statistics, in our typical experiment, we do not address individual atoms; we rather collectively measure the total number of atoms in the state $\ket{m_F}$. Due to the Cs atoms being uncorrelated and identically prepared, the statistics of such a collective measurement can be simply expressed in terms of $q(1|\theta;\lambda)$. Specifically, the probe atoms within each experimental realization are independent and identically distributed (i.i.d.). To this end, we denote $p(n|\theta;\lambda)$ as the probability of detecting $n$ atoms  conditioned on the temperature value being $\theta$ and control parameters being set to $\lambda = (\tau,m_F)$. Note that $p(n|\theta;\lambda) = \sum_{N_0=0}^{\infty} p(n,N_0|\theta;\lambda)$, where $N_0$ is the total number of atoms which itself is a random variable following a Poisson distribution $\mathcal{P}_\mu(N_0)=\mu^{N_0}/N_0!\:\exp(-\mu)$. The mean total probe atom number $\mu$ is independently measured by imaging all probe atoms in the absence of the bath and is thus treated as a fixed parameter in the estimation protocol. One can then use the law of total probability to write
\begin{align}\label{eq:model}
	p(n|\theta;\lambda) &= \sum_{N_0=0}^{\infty} \mathcal{P}_\mu(N_0)\ \mathcal{B}\left[n, N_0,q(1|\theta;\lambda) \right], \\ \nonumber
	&= \mathcal{P}_{\mu q(1|\theta;\lambda)}(n) \\ \nonumber
\end{align}
with the binomial distribution given by $\mathcal{B}\left(n, N_0,q \right)=\binom{N_0}{n}q^n\left(1-q\right)^{N_0-n}$. Here,
we have used that $p(n| N_0,\theta;\lambda)$, the probability of detecting $n$ atoms conditioned on the total number of atoms being $N_0$, the temperature $\theta$, and control parameters $\lambda = (\tau , m_F)$, is given by the Binomial distribution, i.e., $p(n| N_0,\theta;\lambda) = \mathcal{B}\left[n, N_0,q(1|\theta;\lambda) \right]$. It can be shown that this product reduces to a Poisson distribution $\mathcal{P}_{\mu q(1|\theta;\lambda)}(n)$ with mean $\mu q(1|\theta;\lambda)$ \cite{Glatthard2022Optimal}. \\

We emphasize that our atom detection protocol does not yield a continuous classical signal, but a discrete projection onto a quantum state at a given time. The smooth population dynamics shown in \autoref{fig:SE_comparison} arise from averaging over many single-atom spin-exchange trajectories and repeated measurements. At the single-atom level, the dynamics is described by a Markov jump process, namely \autoref{eq:rate}, which fully characterizes ${\hat \rho(\theta;\tau)}$ as the system remains diagonal. The inset of \autoref{fig:SE_comparison}(a) shows a histogram of individual measurements at fixed parameters, illustrating the stochastic nature of the atom-number distribution. Accordingly, since the probe is a quantum system, specifically an ensemble of single atoms with quantized internal spin states, and the measurement outcomes are discrete occupations of these states, we refer to it as a quantum probe \cite{Quantum_sensing_review}.

Equation~\ref{eq:model} efficiently formalizes our model and will be at the heart of our estimation task, as we will discuss below. The temperature dependence of spin populations during the non-equilibrium relaxation constitutes the basis of our atomic quantum probe. After an interaction time $\tau$, the statistics of the observed number of probe atoms in state $\ket{m_F}$ convey information about the bath temperature. For example, a smaller atom number in state $\ket{+3}$ at a given interaction time corresponds to a lower bath temperature, due to faster depopulating spin-exchange dynamics. This behavior is schematically illustrated in \autoref{fig:conceptual_figure}(d). While this is an intuitive argument, one can always quantify the amount of information that one acquires from a specific set of control parameters $\lambda=\left(\tau,m_F\right)$, which then can be used to optimize those.

\subsubsection{Bayesian inference for thermometry}
The Bayesian approach to thermometry is the appropriate choice for scenarios with prior knowledge about the temperature. It is well suited for single shot and finite shots alike, and can be used to examine the transition to the asymptotic limit of large data. In Bayesian estimation theory, one starts with a prior distribution $p(\theta)\geq 0$ satisfying $\int d\theta\,p(\theta) = 1$. This prior represents all the available knowledge about the unknown parameter. One then aims to refine their knowledge by incorporating new data (evidence) about the parameter value. In a quantum gas experiment setup like ours, data is obtained by performing measurements on the probe and reading their outcomes. The task of updating one's knowledge based on the new data amounts to finding the so-called \textit{posterior distribution} $p(\theta| n;\lambda)$, the probability of the parameter being $\theta$ conditioned on observing outcome $n$ and having set the control parameters to $\lambda$. Again, in our specific setup, the control parameters are the interaction time $\tau$ and the measurement choice $m_F$.  To this aim, one uses Bayes' theorem, 
\begin{equation}
	\label{eq:bayes}
	p(\theta| n;\lambda) = \frac{p(n|\theta;\lambda)\,p(\theta)}{p(n|\lambda)}.
\end{equation}
Here, the main component summarizing the probing model is the likelihood function $p(n|\theta;\lambda)$ which is the probability of obtaining the outcome $n$ conditioned on  the parameter taking the value $\theta$ and setting the control parameters to $\lambda=(\tau,m_F)$. For our setup, this was already obtained in \autoref{eq:model}. Furthermore, $p(n|\lambda) = \int d\theta\, p(n|\theta;\lambda)p(\theta)$ is a normalization factor sometimes referred to as the marginal likelihood or evidence. This Bayesian updating procedure is visualized in \autoref{fig:conceptual_figure}(a). 

In this work we consider repeated measurements, which require the evaluation of this update procedure iteratively, allowing us to refine our knowledge about the parameter as we gain more information. Namely, we would like to obtain $p(\theta|{\bm n}\measidx{k};{\bm \lambda}\measidx{k})$---the posterior distribution when having chosen the sequence of control parameters $ \{\lambda\measidx{k}, \lambda\measidx{k-1},\dots \lambda\measidx{1}\} \eqqcolon {\bm \lambda}\measidx{k}$, and after observing the sequence of outcomes $\{n\measidx{k}, n\measidx{k-1}, \dots, n\measidx{1}\} \eqqcolon {\bm n}\measidx{k}$. To this aim, one takes the posterior of the previous round as the prior of the current round. That is, $p\measidx{k}(\theta) \leftarrow p(\theta|{\bm n}\measidx{k-1};{\bm \lambda}\measidx{k-1})$ with the convention that 
$p(\theta|{\bm n}\measidx{0};{\bm \lambda}\measidx{0}) \coloneqq p\measidx{1}(\theta)$ is the initial prior which we use for the first measurement round. With this notation, we have
\begin{align}\label{eq:bayes_kth_shot}
	p(\theta|{\bm n}\measidx{k};{\bm \lambda}\measidx{k})
	& = \frac{p({\bm n_k}|\theta;{\bm \lambda_k})p_1(\theta)}{p({\bm n}_k|{\bm \lambda}_k)}
	\nonumber\\
	&=\prod_{i=1}^{k}\frac{p(n_i|\theta;{\lambda}_i)}{p(n_i|{\bm n}_{i-1};{\bm\lambda}_i)}p_1(\theta),
\end{align}
where we used that $p({\bm n_k}|\theta;{\bm \lambda_k}) = \prod_{i=1}^{k} p(n_i|\theta;{\lambda}_i)$.
Note that the measurement outcomes are not explicitly dependent on the previous outcomes. Thus, the likelihood $ p(n_i|\theta;{\lambda}_i)$ is always described by \autoref{eq:model} as the physical model remains fixed throughout the experiment. Nonetheless, due to day-to-day fluctuations in the experimental setup, some parameters within the model, such as $\mu$ and $N_\text{b}$, may vary. These parameters will be measured prior to the estimation protocol. Furthermore, we have also used the extended definition of the marginal likelihood
\begin{align}
	p(n_i|{\bm n}_{i-1};{\bm\lambda}_i) = \int d\theta\, p(\theta|{\bm n}_{i-1};{\bm \lambda}_{i-1}) p(n_i|\theta; \lambda_i),
\end{align}
and it holds that $p({\bm n}_k|{\bm \lambda}_k) = \prod_{i=1}^k p(n_i|{\bm n}_{i-1};{\bm\lambda}_i)$.

While the posterior represents our knowledge about the parameter probabilistically, often one uses an \textit{estimator} ${\tilde \theta}(\bm n\measidx{k};\bm\lambda\measidx{k})$ to assign a specific value to the parameter, based on the observed data (and of course the prior knowledge). The quality of the estimator can be measured by some error quantifier. A standard error quantifier in quantum thermometry is the squared logarithmic error~\cite{GlobalThermometry,bayesian_estimation_for_collisional_thermometry}, defined as
\begin{equation}
	\begin{aligned}
		\epsilon_\text{log}(\bm n\measidx{k};\bm\lambda\measidx{k}) = \hspace{-.2cm}
		\int \hspace{-.1cm} d\theta\, &p(\theta| {\bm n}\measidx{k};{\bm \lambda}\measidx{k})\log^2\hspace{-.1cm}\left[\frac{\tilde{\theta}(\bm n\measidx{k};\bm\lambda\measidx{k})}{\theta}\right].
	\end{aligned}
	\label{eq:mle}
\end{equation}
The logarithmic loss arises naturally because thermometry is a scale-estimation
problem: overestimating and underestimating the temperature by the same factor are
penalized equally~\cite{GlobalThermometry}. For narrow posteriors,
$\log(\tilde{\vartheta}/\theta)\simeq(\tilde{\vartheta}-\theta)/\theta$, and
\autoref{eq:mle} reduces to the relative mean squared error~\cite{GlobalThermometry}.  

Having quantified the error, one seeks to minimize it. A first minimization comes at the level of data analysis, i.e., over the estimator. The estimator minimizing~\autoref{eq:mle} is analytically known to be~\cite{GlobalThermometry}
\begin{equation}
	\label{eq:estimator}
		\tilde{\vartheta}(\bm n\measidx{k};\bm\lambda\measidx{k})=
		\theta_\text{u}\exp
		\!\left( \int d\theta\, p(\theta| {\bm n}\measidx{k};{\bm \lambda}\measidx{k})\log\left[\frac{\theta}{\theta_\text{u}}\right] \right),
\end{equation}
with $\theta_\text{u}$ an arbitrary unit-eliminating temperature that makes the logarithm dimensionless. However, its value does not affect the estimate $\tilde{\vartheta}$. 
As the optimal estimator is provided by this simple expression, we use it in our entire analysis.
Particularly, inserting \autoref{eq:estimator} into \autoref{eq:mle}, the minimal error is the posterior variance of
$\log\theta$, which for narrow posteriors is set by the posterior variance of  $\theta$~\cite{GlobalThermometry}:
\begin{equation}
	\epsilon_\text{log}(\bm n\measidx{k};\bm\lambda\measidx{k}) = {\rm Var }_{p(\theta|\bm n_k;\bm \lambda_k)}[\log\theta]\simeq\frac{\sigma^2}{[{\tilde \vartheta}(\bm n\measidx{k};\bm\lambda\measidx{k})]^2},
	\label{eq:mle_asympt}
\end{equation}
with $\sigma^2 \coloneqq {\rm Var }_{p(\theta|\bm n_k;\bm \lambda_k)}[\theta]$.
This motivates the absolute uncertainty
$\Delta\tilde{\vartheta} \equiv \tilde{\vartheta}\sqrt{\epsilon_\mathrm{log}} \simeq \sigma$,
which has units of temperature and coincides with the posterior standard deviation
for narrow posteriors~\cite{Glatthard2022Optimal}. In Sec.~\ref{sec:results} we use $\Delta\tilde \vartheta$ to illustrate error bars on the estimates.
\subsubsection{Bayesian information gain}
\label{subsubsec:info_gain}
Let us denote $\bar \epsilon_{\text{log}}({\bm n}\measidx{k-1};{\bm \lambda}\measidx{k})$ as the mean of the error over all possible outcomes of the $k$th measurement, given a previously observed trajectory of outcomes ${\bm n}\measidx{k-1}$, that is
\begin{align}\label{eq:error_inf_gain}
	{\bar \epsilon}_{\rm log}({\bm n}\measidx{k-1};{\bm \lambda}\measidx{k}) & = \sum_{n\measidx{k}} p(n\measidx{k}|\lambda\measidx{k}) { \epsilon}_{\rm log} ({\bm n}\measidx{k};{\bm \lambda}\measidx{k}) \nonumber\\
	& = \epsilon_\text{log}({\bm n}\measidx{k-1};{\bm \lambda}\measidx{k-1}) - {\cal K}({\bm n}\measidx{k-1};{\bm \lambda}\measidx{k}),
\end{align}
where ${\cal K}({\bm n}\measidx{k-1};{\bm \lambda}\measidx{k})$ is the expected \textit{information gain} attributable to the $k$th measurement, defined as~\cite{GlobalThermometry}
\begin{align}
	\label{eq:K}
	{\cal K}(\bm n\measidx{k-1};\bm\lambda\measidx{k}) =\sum_{n\measidx{k}} p(n\measidx{k}|\lambda\measidx{k}) \log^2\left[\frac{\tilde{\vartheta}(\bm n\measidx{k};\bm\lambda\measidx{k})}{\tilde{\vartheta}(\bm n\measidx{k-1};\bm\lambda\measidx{k-1})}\right]. 
\end{align} 
The information gain is a central object in our analysis, particularly in the design of optimal strategies, and thus it is worth exploring it further here. From \autoref{eq:error_inf_gain} we note that the first term, namely $\epsilon_\text{log}({\bm n}\measidx{k-1};{\bm \lambda}\measidx{k-1})$, entirely depends on the current prior distribution (and through that on all the previous values for the control parameters $\bm \lambda_{k-1}$). The expected information gain $\mathcal{K}$ is thus the only contribution that the current control parameters $\lambda_k$ can influence. One can thus choose the optimal value for the control parameters by maximizing the expected information gain, which is the basis of two of our strategies, as discussed in the next section.

To motivate the importance of tuning the control parameters appropriately, note the following limiting cases: a measurement at interaction time $\tau=0$ yields no temperature information, as all trajectories for different bath temperatures $\theta$ share the same initial state. Analogously, for large $\tau$, all trajectories evolve towards polarization in the state $m_F=-3$, again resulting in a common state, independent of the bath temperatures $\theta$, and therefore yielding a vanishing information gain. Consequently, between these two limits, projection on each state $\ket{m_F}$ must exhibit an optimal interaction time at which the extracted information is maximal.
These optimal interaction times depend on the prior temperature distribution, making the problem particularly well suited for adaptive strategies where one would tune the parameters as the prior gets updated. This dependence on the temperature distribution is rooted in the scaling of the spin-exchange evolution with temperature. When lower temperatures are more probable under the current prior, the optimization favors shorter interaction times, because lower temperatures lead to faster spin-exchange evolution.

\begin{figure}[t]
	\subfloat{\includegraphics[]{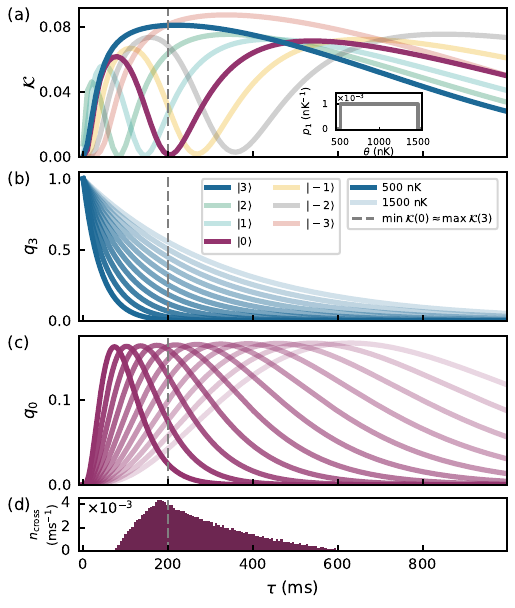}}\\
	\caption{\textbf{Optimal measurement parameters.} \\(a) Calculated mean information gain $\mathcal{K}$ (\autoref{eq:K}) for a measurement performed at interaction time $\tau$ (abscissa) for different measured states $\ket{m_F}$ (colors), assuming no prior knowledge, i.e., a uniform prior probability distribution function (inset). The states $\ket{+3}$ (blue) and $\ket{0}$ (purple) are highlighted as representative cases. While the edge states ($\ket{\pm3}$) exhibit unimodal behavior, the intermediate states ($\ket{2}$ to $\ket{-2}$) show bimodal time dependence with a pronounced minimum of nearly vanishing information gain.
		(b), (c) Calculated spin-exchange-induced population dynamics $q_{m_F}$ for (b) $m_F=3$ and (c) $m_F=0$ for bath temperatures $T$ between $\SI{500}{\nano\kelvin}$ and $\SI{1500}{\nano\kelvin}$. Higher opacities correspond to lower temperatures $\theta$. The dynamics for $\ket{+3}$ are clearly separated, whereas those for $\ket{0}$ exhibit intersections. (d) Histogram $n_\mathrm{cross}$ of the intersection times of the trajectories $q_0(\theta;\tau)$ for state $\ket{0}$, as a function of the interaction time $\tau$. Intersections are counted between all pairs of hypothetical temperatures $\theta_1$ and $\theta_2$, each chosen from $N = 1000$ values uniformly distributed between the minimum and maximum bath temperatures. A high number of intersections coincides with a low mean information gain $\mathcal{K}$ (vertical gray dashed line).}
	\label{fig:K_and_two_mF}
\end{figure}

We now demonstrate the dependence of the expected information gain on the control parameters for a simple single-measurement scenario. Consequently, there is no dependence on the previous outcomes, hence, we denote the mean information gain simply by ${\cal K}(\tau,m_F)$. We choose a flat initial prior on the interval $[\theta_\text{min},\theta_\text{max}]=[\SI{500}{\nano\kelvin},\SI{1500}{\nano\kelvin}]$ (see the inset of \autoref{fig:K_and_two_mF}(a)) for its simplicity and as a pragmatic representation of the absence of prior knowledge. This choice is sufficient because the experimentally feasible probing interval spans less than one order of magnitude. Hence, a symmetry- or geometry-informed prior \cite{Nimmrichter_Rubio_symmetry_geometry}, such as Jeffreys' prior \cite{Boeyens2021Uninformed}, only changes the estimates by a few percent at the beginning of the measurement sequence and has no effect in the many-measurement regime or on the posterior convergence. Moreover, as discussed later in Sec.~\ref{subsec:fisher_info}, the posterior rapidly converges toward a Gaussian distribution and becomes insensitive to the initial prior after the first few measurements. For larger temperature probing intervals, the initial choice of the prior is more influential in the analysis. Thus, one should exercise great care in choosing the prior that reflects appropriately their ignorance or knowledge about the temperature.

As shown in \autoref{fig:K_and_two_mF}(a), $\mathcal{K}(\tau, m_F)$ depends nontrivially on both control parameters. Firstly, note that the shape of $\mathcal{K}(\tau, m_F)$ differs between the edge states ($m_F \in \{-3,+3\}$) and the intermediate states ($m_F \in \{-2,-1,0,+1,+2\}$). For the edge states, it is unimodal. For the intermediate states, it is bimodal, with a pronounced minimum of nearly vanishing information gain. For each state, the information gain is maximal at a finite interaction time, as expected from the limiting cases above. For our flat prior, the largest value of information gain is reached for the edge state $\ket{+3}$, which therefore sets the parameters of the a~priori optimized protocol, see Sec. \ref{subsubsec:apo_protocol}.

The difference in behavior between the edge and intermediate states originates from the population dynamics. The population of $\ket{+3}$ ($\ket{-3}$) decreases (increases) monotonically. Trajectories for different temperatures therefore do not cross, see \autoref{fig:K_and_two_mF}(b). The intermediate states start empty, reach a maximum, and decay back to zero. Their trajectories for different temperatures therefore intersect, see \autoref{fig:K_and_two_mF}(c).  At an intersection, two hypothetical temperatures $\theta_1$ and $\theta_2$ yield the same expected population. Within our model, they thus yield the same likelihood, $p(n|\theta_1;\tau,m_F) = p(n|\theta_2;\tau,m_F)$. Even a perfect measurement cannot distinguish whether a given population stems from the rising branch of a hot-bath trajectory or from the falling branch of a cold-bath trajectory. This ambiguity suppresses the information gain.

To support this picture, we compute the times at which the trajectories $q_0(\theta;\tau)$ for different temperatures intersect. We sample $N = 1000$ temperatures uniformly between the minimum and maximum bath temperatures and record the intersection time for each pair. As shown in \autoref{fig:K_and_two_mF}(d), the minimum of $\mathcal{K}(\tau,m_F=0)$ (gray dashed line) lies close to the maximum of the histogram of intersections. At the same interaction time, the trajectories of $\ket{+3}$ are separated, and $\mathcal{K}$ for $\ket{+3}$ is close to its maximum. Measurements at parameters with many intersections, e.g., $\lambda = (\tau = 200\,\mathrm{ms}, m_F = 0)$, can lead to a bimodal posterior, see Sec.~\ref{subsec:bimodal_posterior} and \autoref{fig:splitting_posterior} in the Supplemental Material.

\subsection{Measurement Protocols}
\label{subsec:measuement_protocols}
We realize three different measurement protocols and use two inference methods to evaluate the acquired data. The measurement protocols characterize the sequence of measurement parameters, while the inference methods define how temperature estimates are extracted from the acquired data. The three measurement protocols are the unoptimized, a~priori optimized, and adaptively optimized protocols. As the names suggest, they differ in the degree of optimization used to determine (optimal) measurement parameters. 

\subsubsection{Unoptimized Protocol}
\label{subsubsec:uo_protocol}
A simple approach, which we call the \textit{unoptimized} protocol, is to ignore
the dependence of the information gain on the control parameters, and instead sample the full parameter range uniformly. Particularly, we sample from the parameter vector
\begin{equation}
	\boldsymbol{\lambda}^{\mathrm{uo}}_k=\left\{[\tau\measidx{k},(m_{F})_k],\dots,[\tau\measidx{2},(m_{F})_2],[\tau\measidx{1},(m_{F})_1]\right\},
	\label{eq:parameters_uo}
\end{equation}
where for the interaction time $\tau$ we choose 20 equidistant values between $\SI{0}{\milli\second}$ and $\SI{1200}{\milli\second}$. This range is sufficient to capture the dynamics over all relevant evolution speeds. We perform measurements for this protocol by sequentially scanning over these parameters from short to long interaction times and then from the initial state $\ket{+3}$ to the final state $\ket{-3}$. To eliminate the influence of this naive parameter ordering, all datasets are evaluated by Bayesian inference after post-processing the measurements into a randomized order that is identical for all datasets. This random order is indicated for the interaction time $\tau$ in the left inset of \autoref{fig:Comparison}(a). Due to the changing measurement parameters, the individual measurement outcomes are not identically distributed, $p(n_k|\theta;\lambda_k)\neq p(n_{k'}|\theta;\lambda_{k'})$. However, they remain statistically independent because the parameter sequence $\bm\lambda_k^\mathrm{uo}$ is predetermined and each measurement is performed on an independently prepared experimental realization.

\subsubsection{A~priori optimized Protocol}
\label{subsubsec:apo_protocol}
Calculating the optimal parameters only once constitutes the \textit{a~priori optimized} protocol,
\begin{equation}
	\boldsymbol{\lambda}^{\mathrm{apo}}_k=\left\{[\accentset{\ast}{\tau}\measidx{1},(\accentset{\ast}{m}_{F})_1],\dots,[\accentset{\ast}{\tau}\measidx{1},(\accentset{\ast}{m}_{F})_1],[\accentset{\ast}{\tau}\measidx{1},(\accentset{\ast}{m}_{F})_1]\right\},
	\label{eq:parameters_apo}
\end{equation}
with 
\begin{align}\label{eq:opt_apo}
	[\accentset{\ast}{\tau}\measidx{1},(\accentset{\ast}{m}_{F})_1] \coloneqq \underset{[\tau\measidx{1},(m_{F})_1]}{{\rm argmax}}\,{\cal K}({\bm \lambda}\measidx{1}).
\end{align}
That is, all measurements of each dataset in the a~priori optimized protocol are performed at those parameters that maximize the information gain for the first round. See also the left inset of \autoref{fig:Comparison}(b). Since these parameters are kept fixed throughout the dataset, the experimental outcomes are independent and identically distributed (i.i.d.). For our specific choice of prior, it turns out that $\ket{(\accentset{\ast}{m}_{F})_1} = \ket{+3}$, and $\accentset{\ast}{\tau}\measidx{1}$ varies slightly from $\SI{200}{\milli\second}$ to $\SI{280}{\milli\second}$, due to day-to-day fluctuations in the bath atom number and the exact initial spin state of the probe atoms.

\subsubsection{Adaptive Protocol}
\label{subsubsec:ao_protocol}
A straightforward implementation of Bayesian feedback is the greedy adaptive protocol. At each step $k$, this strategy tunes the control parameters to those that maximize the subsequent information gain ${\cal K}({\bm n}_{k-1};{\bm \lambda}\measidx{k})$, thereby implementing real-time feedback. However, in our setup, we fix the $m_F$ parameter only once to reduce the computational effort required for the adaptive optimization, but we keep updating the interaction time. Mathematically speaking, we have

\begin{equation}
	\boldsymbol{\lambda}^{\mathrm{ado}}_k=\left\{[\accentset{\ast}{\tau}\measidx{k},(\accentset{\ast}{m}_{F})_1],\dots,[\accentset{\ast}{\tau}\measidx{2},(\accentset{\ast}{m}_{F})_1],[\accentset{\ast}{\tau}\measidx{1},(\accentset{\ast}{m}_{F})_1]\right\},
	\label{eq:parameters_ao}
\end{equation}
with $[\accentset{\ast}{\tau}\measidx{1},(\accentset{\ast}{m}_{F})_1]$ given by \autoref{eq:opt_apo}, and
\begin{align}
	\accentset{\ast}{\tau}\measidx{k} \coloneqq \underset{\tau\measidx{k}}{{\rm argmax}}\,{\cal K}({\bm n}_{k-1};\{[\tau_k,(\accentset{\ast}{m}_F)_1],{\bm \lambda}\measidx{k-1}\}),~~k>1.
\end{align}
As the strategy looks only one step ahead it is not necessarily globally optimal. Nonetheless, it can significantly improve the precision and is often asymptotically optimal for $k\gg 1$ \footnote{An intuitive way to see asymptotic optimality is via the lens of the Bernstein-von Mises theorem. When the latter holds, the posterior becomes very sharp around the true parameter value and stops changing significantly. The optimal parameters therefore converge to those suitable for the true parameter, which hence cannot be improved.}. The feedback, namely the dependence of the measurement parameter $\lambda_k$ on the previous outcome record $\bm n_{k-1}$ and parameter record $\bm \lambda_{k-1}$, makes the outcome sequence history dependent. Consequently, the outcomes are neither independent nor identically distributed. Asymptotically, however, the measurement parameters stabilize around the optimum selected by the converged posterior, so that the later measurements become approximately i.i.d.

In the adaptive strategy, the optimal interaction time takes a different value for each measurement, typically ranging from $\SI{100}{\milli\second}$ when the current estimate of the bath temperature favors the colder end of the prior domain, up to $\SI{700}{\milli\second}$ for the hotter end. An exemplary adaptive change of the interaction time is shown in the left inset of \autoref{fig:Comparison}(c).

Because of a low experimental repetition rate, caused by the adaptive calculations and changing experimental settings after each measurement, the bath atom number $N_\mathrm{b}$ drifts slightly, while the true temperature $T$ remains constant within the precision of the time-of-flight thermometry. Thus, the likelihood function needs to be adapted frequently. To achieve this, we perform independent TOF measurements to determine $N_\mathrm{b}$ every 50 probe measurements and adapt the likelihood function accordingly. \\

While the focus of this work and our primary inference method used for all protocols is the Bayesian approach, for the unoptimized protocol only, we also perform a frequentist $\chi^2$-based fitting procedure to determine the bath temperature from the best-fitting rate-equation simulation \cite{chi_squared_ref, SpinThermometry}.

	\section{Results}
\label{sec:results}

We now compare the previously introduced protocols using multiple criteria. First, we analyze the convergence of the Bayesian estimate, together with the corresponding evolution of the logarithmic error and the mean information gain during the estimation process. We then quantify the accuracy and temperature dependence of the final estimates with respect to reference TOF thermometry. The convergence speed is compared through the number of measurements required to reach a given target precision. Finally, we investigate the asymptotic behavior of the different protocols. For this, we compare their mean estimation errors with the expected Cramér--Rao bound (ECRB) and relate the posterior variance to the accumulated Fisher information as predicted by the Bernstein--von Mises theorem.

\subsection{Estimation Convergence and Information Gain}
\label{subsec:estimation_convergence}
\begin{figure*}[p]
	\subfloat{\includegraphics[]{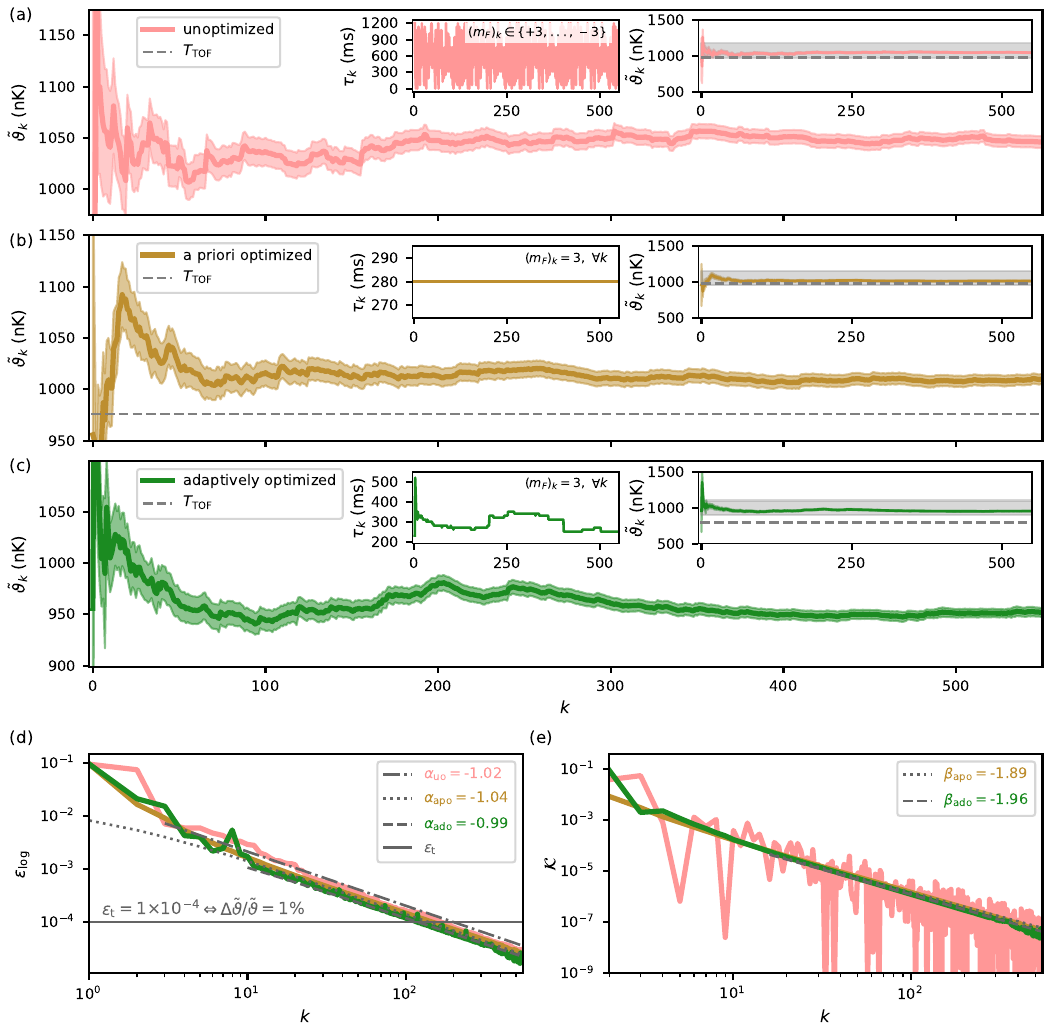}}
	\caption{\textbf{Evolution of estimates $\tilde\vartheta$, mean logarithmic error $\epsilon_\mathrm{log}$, and mean information gain $\mathcal{K}$.}\\ (a)–(c) Convergence of the temperature estimates $\tilde\vartheta$ for the (a) unoptimized, (b) a~priori optimized, and (c) adaptively optimized protocols. The measurement index is given by $k$. The shaded areas indicate the absolute estimation errors defined as $\Delta\tilde\vartheta=\tilde{\vartheta}\sqrt{\epsilon_\text{log}}$. Right insets show the same data over the full temperature range; the gray shaded areas correspond to the regions shown in the main panels. The dashed gray lines mark the temperatures determined from independent time-of-flight (TOF) reference measurements. Left insets show the interaction time parameter $\tau$ for the three protocols: randomized scan, constant, and adaptively varied. (d) Evolution of the mean logarithmic error $\epsilon_\text{log}$ for all three strategies. Dash-dotted, dotted, and dashed lines show power-law regressions of the form $\epsilon_\text{log}\propto k^\alpha$, starting after three measurements for the unoptimized protocol, after the first one for the a~priori protocol and after 10 measurements for the adaptive protocol. Extracted exponents $\alpha$ are given in the legend. The solid horizontal gray line marks the threshold $\epsilon_\text{t}=1\times10^{-4}$ used to determine the threshold measurement number $k_\mathrm{t}$. (e) Evolution of the mean information gain $\mathcal{K}$. Dotted and dashed lines show power-law regressions of the form $\mathcal{K}\propto k^\beta$, starting after 15 measurements. Extracted exponents $\beta$ are given in the legend. No regression is performed for the unoptimized protocol, as it does not exhibit power-law scaling.}
	\label{fig:Comparison}
\end{figure*}

The iterative application of Bayes’ theorem uses the collected data to adjust the posterior by shifting its mean and narrowing down its width.
This causes the temperature estimate to converge toward a final value, as illustrated in \autoref{fig:Comparison}(a)-(c) for one exemplary measurement sequence of each protocol. During the first few measurements, the estimate changes rapidly, reflecting the high mean information gain $\mathcal{K}$, resulting from the limited prior knowledge. This is clearly visible in the right insets of \autoref{fig:Comparison}(a)-(c), where the same data as in the main panel is plotted over the full probing range. After a few hundred measurements, sufficient information has been obtained to yield a reasonable estimate, as the posterior has narrowed substantially around the estimated value. The final estimate exhibits a small offset toward higher temperatures relative to the temperature obtained from time-of-flight (TOF) thermometry, which is discussed in Sec.~\ref{subsec:accuracy}. A dataset of the unoptimized protocol converges to the same final estimate, irrespective of the order of the measurement parameters, but requires different numbers of measurements to become reliable.
\newline
As a consequence of the posterior narrowing, the estimation error $\epsilon_\text{log}$ defined in \autoref{eq:mle} also decreases. This can be seen in \autoref{fig:Comparison}(d), and in the reduction of the absolute error $\Delta\tilde{\vartheta} = \tilde{\vartheta}\sqrt{\epsilon_\mathrm{log}}$, indicated by shaded areas around the estimate. After an initial unstable phase of a few tens of measurements, the estimation error exhibits a power-law scaling with the number of measurements, $\epsilon_\text{log}\propto k^\alpha$. The fitted exponents of these power laws are $\alpha\approx\num{-1}$ for all three protocols. In Sec.~\ref{subsec:fisher_info} we come back to this point, where we show this inverse scaling with $k$ for all strategies subject to large $k$. 

As quantified later in Sec.~\ref{subsec:conv_speed}, and as is already apparent from these exemplary datasets, the estimation error is smallest for the adaptive protocol, followed by the a~priori optimized protocol, and finally the unoptimized protocol. Transient deviations from this ordering can occur for individual datasets, since experimental fluctuations, for example in the total atom number, can produce measurement outcomes that are temporarily inconsistent with the current posterior. In such cases, the posterior may narrow more slowly or even broaden over a few measurements before the asymptotic power-law scaling is recovered.

The power-law reduction of the estimation error is reflected in the decay of the mean information gain $\mathcal{K}$, shown in \autoref{fig:Comparison}(e). The two optimized protocols exhibit a power-law scaling, $\mathcal{K}\propto k^\beta$, with fitted exponents of $\beta_\text{apo}\approx\beta_\text{ado}\approx\num{-2}$. By contrast, the unoptimized protocol shows pronounced minima in the acquired information. These correspond to the minima discussed in connection with \autoref{fig:K_and_two_mF}(a) and arise because the unoptimized protocol repeatedly samples measurement parameters that yield minimal information.
The observed ${\mathcal{K}\propto k^{-2}}$ scaling follows naturally from the relation between the mean information gain and the expected reduction of the mean logarithmic error. The mean information gain $\mathcal{K}$ is the decrease of $\bar\epsilon_\text{log}$ per additional measurement, $\mathcal{K}_k=\epsilon_{k-1}-\bar\epsilon_k$ \cite{GlobalThermometry}. Since $\epsilon_k\propto k^{-1}$, this gives \mbox{$\mathcal{K}_k = \epsilon_{k-1}-\bar\epsilon_{k} \propto (\frac{1}{k-1}-\frac{1}{k}) = \frac{1}{k(k-1)} \sim k^{-2}$.}

In conclusion, all protocols lead to convergent temperature estimates, with additional measurements continuing to reduce the uncertainty. The observed power-law reduction of the mean logarithmic estimation error follows $\epsilon_\mathrm{log}\propto 1/k$.

\subsection{Comparison with Reference Thermometry}
\label{subsec:accuracy}

Having established the convergence behavior, we now assess the accuracy of the final estimates $\tilde\vartheta_\mathrm{f}$ by comparing them with the independent time-of-flight (TOF) temperature $T_\text{TOF}$ in \autoref{fig:T_SE_vs_TOF}. The TOF measurement and analysis are described in Sec.~\ref{sec:TOF} of the Supplemental Material. Estimates are considered final after 700 measurements for the unoptimized protocol, corresponding to five repetitions of each parameter tuple \mbox{$\lambda=(\tau, m_F)$}. For the two optimized protocols, we perform between 200 and 800 measurements to obtain a final estimate $\tilde\vartheta_\mathrm{f}$. Each dataset contains enough measurements for the individual estimation error to fall below the threshold defined in Sec.~\ref{subsec:conv_speed} and indicated in \autoref{fig:Comparison}(d). Within the uncertainties attributed to the TOF reference measurements, all estimates of a given protocol lie on a straight line parallel to the identity $\tilde{\vartheta}_\mathrm{f}=T_\text{TOF}$, with small protocol-dependent offsets toward higher temperatures. These offsets are not temperature dependent. For the unoptimized protocol, we show both the Bayesian analysis and the frequentist $\chi^2$ fit to the averaged full spin dynamics. The fitting procedure is described in
Sec.~\ref{subsec:chi_squared} of the Supplemental Material, with an example shown in \autoref{fig:chi_squared}.
\begin{figure}[t]
	\subfloat{\includegraphics[]{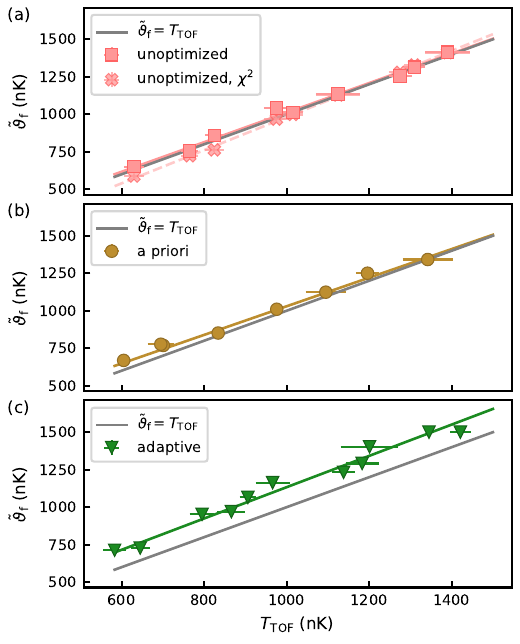}}\\
	\caption{\textbf{Evaluation of final estimates.} Comparison of the temperature measured with the spin probe $\tilde\vartheta_\mathrm{f}$ (ordinate) and the time-of-flight (TOF) reference $T_\text{TOF}$ (abscissa) for (a) unoptimized, (b) a~priori optimized, and (c) adaptively optimized protocols. The identity $\tilde\vartheta_\mathrm{f}=T_\text{TOF}$ is shown as a gray line. Unoptimized measurements (a) are analyzed using two methods: a Bayesian approach (squares) and a $\chi^2$ analysis (crosses). Colored lines show linear regressions to the data. Solid lines correspond to the Bayesian approach, while the dashed line in (a) corresponds to the $\chi^2$ analysis.}
	\label{fig:T_SE_vs_TOF}
\end{figure}
The accuracy, the agreement between the spin-probe and TOF estimates, is good for all protocols except the adaptive one, which shows an average bias toward larger temperatures of approximately $\SI{136\pm12}{\nano\kelvin}$. This offset is likely associated with its slower experimental repetition rate due to the additional time required for the adaptive calculations. Since TOF thermometry is performed at the standard high repetition rate, the slower adaptive acquisition can reduce the comparability of the spin-probe estimates and the TOF reference thermometry. We emphasize that this discrepancy does not reflect a limitation of the adaptive inference strategy itself, but rather the additional experimental overhead introduced by real-time parameter updates. As this bias is approximately temperature independent, such a systematic offset could be easily corrected by calibration.

The similar estimates obtained from the Bayesian (red squares) and frequentist (red crosses) analyses for the unoptimized protocol indicate that both inference methods yield consistent final estimates in the many-measurement limit. The slope obtained from the frequentist $\chi^2$ analysis, however, indicates a more pronounced temperature dependence of the deviation from the TOF reference. This behavior originates from the interaction-time sampling chosen for the unoptimized protocol: 20 equidistant values between $\SI{0}{\milli\second}$ and $\SI{1200}{\milli\second}$. This range provides good sampling of the slowest SE dynamics, which occur at the highest bath temperature of the probing interval, $\SI{1500}{\nano\kelvin}$. For cooler baths, however, measurements at longer interaction times interrogate a probe ensemble that is already polarized in state $m_F=-3$, as shown in \autoref{fig:SE_comparison}(b). Thus, the number of informative measurements per full iteration through parameter space is reduced for cold baths, leading to less accurate estimates. This effect is less pronounced in the Bayesian inference. For larger temperature probing intervals, the unoptimized protocol is expected to exhibit a stronger temperature dependence of the accuracy, as the predefined interaction-time sampling could render an even larger fraction of the probing interval uninformative.

In conclusion, all protocols produce estimates that depend approximately linearly on the reference temperature estimate $T_\text{TOF}$, with little temperature dependence of the bias. The most pronounced deviation between spin-probe and TOF temperature estimates occurs for the adaptive protocol, likely caused by the slower adaptive acquisition reducing the comparability of the two thermometry measurements.

\subsection{Error-Threshold Measurement Number}
\label{subsec:conv_speed}
\begin{figure}[t]
	\includegraphics[]{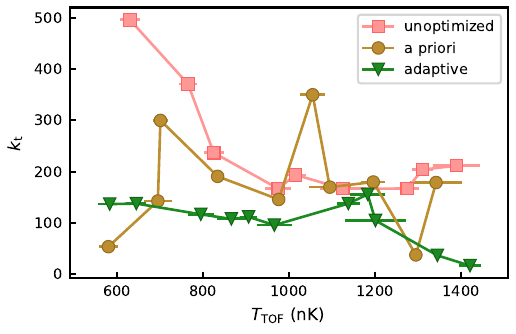}
	\caption{\textbf{Error-threshold measurement number.} Number of measurements $k_\mathrm{t}$ required for the estimation error to fall below the threshold $\epsilon_\text{t}$ marked in \autoref{fig:Comparison}(d), versus the estimated bath temperature $T_\mathrm{TOF}$. Each data point corresponds to one estimation-convergence dataset with a reference temperature $T_\mathrm{TOF}$.}
	\label{fig:convergence_speed_vs_T}
\end{figure}

So far, we discussed the estimation convergence only qualitatively and quantified the final estimates and their accuracy. We now focus on the scarce-data regime, i.e., at the beginning of dataset acquisition, when only a few tens to a few hundred measurements have been taken. We therefore study how quickly the different Bayesian protocols lead to a reliable estimate. To this end, we compare the threshold measurement number $k_\mathrm{t}$, defined as the first measurement number for which the logarithmic error $\epsilon_\mathrm{log}$ falls below the threshold $\epsilon_\mathrm{t}=\num{1e-4}$, corresponding to a relative estimation error of $\Delta\tilde\vartheta/\tilde\vartheta=\SI{1}{\percent}$. This threshold is sufficiently small for all protocols to have reached their asymptotic power-law scaling, see \autoref{fig:Comparison}(d). Figure~\ref{fig:convergence_speed_vs_T} displays the number of measurements required to reach $\epsilon_\mathrm{t}$ for datasets at different temperatures.
Although all protocols show different quantitative dependencies of threshold measurement number $k_\mathrm{t}$ on the estimated bath temperature, a clear hierarchy between the protocols emerges. As already apparent from \autoref{fig:Comparison}(d), the adaptive protocol requires the fewest measurements to reach the threshold precision, followed by the a~priori optimized protocol. On average, the threshold is reached after $\bar k_\mathrm{t,\,ado}=\num{106\pm41}$ and $\bar k_\mathrm{t,\,apo}=\num{173\pm88}$ measurements, with no strong dependence on the bath temperature. However, the situation is different for the unoptimized protocol. Here a clear trend is visible: fewer measurements are required to reach the desired precision when probing a hot bath than when probing a cold one. While the required number of measurements to probe a hot bath is comparable to that of the optimized protocols, about five times more measurements, $k_\mathrm{t,uo}\approx500$, are required for cold bath temperatures. This trend has the same origin as the accuracy trend discussed in Sec.~\ref{subsec:accuracy}: the choice of the interaction-time range favors the slow dynamics, corresponding to hot bath temperatures. The colder the bath, the higher the number of uninformative measurements, see \autoref{fig:SE_comparison}(b).
\newline
The overall ranking of the protocols is expected, since the optimization goal is precisely the minimization of the mean logarithmic error. This makes the adaptive protocol the fastest, followed by the a~priori optimized protocol and finally the unoptimized protocol.
We note that the number of measurements does not account for the laboratory time required to perform them. When laboratory time is used as the resource, the a~priori optimized protocol is favorable, see Sec. \ref{subsec:k_t_in_lab_time} and \autoref{fig:conv_speed_lab_time} in the Supplemental Material.
\subsection{The asymptotic limit of large data}
\label{subsec:fisher_info}
Let us now shift our focus to the asymptotic limit of large data. In this limit (${k\gg 1}$), the Fisher information, a key figure of merit in the frequentist approach, also becomes relevant in the Bayesian approach. We first bound the estimation error in terms of the Fisher information. We then discuss how the Fisher information scales with $\mu$ and $k$, and use it to describe the asymptotic posterior.

\subsubsection{Bounds on the estimation error}
To begin with, let us define
\begin{align}\label{eq:mean_mle}
	\bar\epsilon_{\text{log}}^k = \sum_{{\bm n}_k}p({{\bm n}_k|{\bm \lambda}_k})\epsilon_{\text{log}}({\bm n_k};{\bm \lambda}_k
	),
\end{align}
as the estimation error \textit{on average} after $k$ measurements. The definition is valid for any strategy, namely, those in which the control parameters are a function of previous outcomes $\lambda_i = \lambda_i({\bm n}_{i-1},{\bm \lambda}_{i-1})$, whose dependence we drop here to lighten the notation. For regular priors that vanish at the parameter boundaries, the Van Trees inequality
states that~\cite{van2004detection,Gill1995,Mehboudi2022Fundamental}
\begin{align}\label{eq:vTSB}
	\bar\epsilon_{\text{log}}^k \geq \frac{1}{Q+\int d\theta\;p\measidx{1}(\theta)\;\theta^2\;\mathcal{F}\measidx{k}^\text{acc}\left(\theta\right)}.
\end{align}
Here, $Q$ is the prior information defined as
\begin{align}
	Q=\int d\theta \; p\measidx{1}(\theta) \left[1+\theta\;\partial_\theta\log p\measidx{1}(\theta)\right]^2,
\end{align}
and the accumulated Fisher information is defined as
\begin{align}\label{eq:FI_acc_def}
	\mathcal{F}\measidx{k}^\text{acc}\left(\theta\right)&=\sum_{i=1}^{k}\sum_{\bm n_{i-1}} p({\bm n}_{i-1}|\theta;{\bm\lambda}_{i-1}) \mathcal{F}^\text{ind}(\theta;{\lambda}_{i}),
\end{align}
where $p({\bm n}_{i-1}|\theta;{\bm\lambda}_{i-1})$ is the likelihood of the history of outcomes ${\bm n_{i-1}}$, given the history of control parameters ${\bm \lambda_{i-1}}$ and the hypothesis temperature $\theta$. Here, $\mathcal{F}^\text{ind}$ is the Fisher information of an individual measurement with control parameters $\lambda_i$,
\begin{align}\label{eq:FI_ind_main}
	\mathcal{F}^\text{ind}(\theta;{\lambda}_{i})
	& = \sum_{n_i}p(n_i|\theta;\lambda_i)\left[ \partial_\theta\log p(n_i|\theta;\lambda_i) \right]^2\nonumber\\
	& = \mu\,  q(1|\theta;\lambda_i)[\partial_\theta {\log q(1|\theta;\lambda_i)}]^{2}.
\end{align}
For a proof of the last identity, see Sec. \ref{sec_FI_analatic} in the Supplemental Material.

The Van Trees inequality holds for any estimator, but it is not necessarily tight. Moreover, it requires priors that vanish at the boundaries of the parameter interval. This excludes our flat prior. 
Defining the estimation bias as 
\begin{align}
	b_k(\theta) = \sum_{{\bm n}_k}p({\bm n}_k|\theta;{\bm \lambda}_k)\log {\tilde \vartheta}({\bm n}_k;{\bm \lambda}_k) - \log \theta,
\end{align}
one can use the standard Cram\'er-Rao bound to obtain the expected Cram\'er-Rao bound (ECRB)~\cite{van2004detection,1998},
\begin{align}\label{eq:bcrb_unbiased}
	\bar\epsilon^k_{\text{log}} & \overset{\rm }{\geq} \int d\theta p\measidx{1}(\theta) \left[\frac{[1+\theta\partial_{\theta}b_k(\theta)]^2}{\theta^2 {\cal F}^\text{acc}\measidx{k}(\theta)} + b_k(\theta)^2\right]\nonumber\\
	& \overset{\rm u.b.}{=} \int d\theta  \frac{p\measidx{1}(\theta)}{\theta^2 {\cal F}^\text{acc}\measidx{k}(\theta)}.
\end{align}
Here, the second expression only holds for unbiased estimators. For finite data, unbiased estimators may not exist. For $k \gg 1$, however, reasonable estimators become unbiased, and thus we can use the expression. In this regime, the prior information becomes negligible compared to the accumulated Fisher information, and we expect the bound to be saturable. By Jensen's inequality, it is then tighter than \autoref{eq:vTSB}.
\subsubsection{Scaling of the Fisher information}
Both bounds are set by the accumulated Fisher information. We therefore discuss how it scales with the mean number of probe atoms $\mu$ and with the number of measurements $k$. An immediate consequence of \autoref{eq:FI_ind_main} is the linear scaling with $\mu$, which is a demonstration of the standard quantum limit (SQL). This linear dependence carries over directly to the accumulated Fisher information for all three strategies considered here. Scaling beyond the SQL, up to the Heisenberg limit with ${\cal F}^{\rm ind}(\theta;\lambda)\propto \mu^2$, would require resources such as entanglement between the atomic probes \cite{quantum_metrology_cold_gases_review} which are currently beyond the reach of our experiment.
We will now show that, in the asymptotic limit, we also have a linear scaling with $k$ for all three strategies, something that we qualitatively observed in \autoref{fig:Comparison}(d).

For the unoptimized protocol, the parameters are chosen at the beginning from a mesh, so that the schedule is fixed in advance. In this case, we have $\lambda_i$ fully independent of the outcomes, thus
\begin{align}\label{eq:accumulated_FI_unoptimized}
	\mathcal{F}\measidx{k}^\text{acc}\left(\theta\right)&\overset{\text{uo}}{=}\sum_{i=1}^{k} \mathcal{F}^\text{ind}(\theta;{\lambda_i}) \sum_{\bm n_{i-1}} p({\bm n}_{i-1}|\theta;{\bm\lambda}_{i-1}) \nonumber\\
	&=\sum_{i=1}^k\mathcal F^\text{ind}(\theta;\lambda_i).
\end{align}
Clearly, this is additive in $k$. Note that the quantity, however, depends on the ordering of the mesh. To make the result independent of how the $\lambda_j$ are labeled, we randomly permute the $m$ mesh points and select the first $k$ elements. Because each mesh point $\lambda_j$ has an equal marginal probability $1/m$ of being selected at any step $i$, the expected Fisher information for the randomized setting is
\begin{align}\label{eq:uno_additive_FI}
	\mathcal{F}\measidx{k}^\text{acc}\left(\theta\right)&\overset{\text{ruo}}{=}\sum_{i=1}^k\frac{1}{m}\sum_{j=1}^m\mathcal F^\text{ind}(\theta;\lambda_j) = \frac{k}{m}\sum_{j=1}^m\mathcal F^\text{ind}(\theta;\lambda_j),
\end{align}
which is linear in $k$.

For the a~priori setting, linearity in $k$ is more trivial. Since the control parameters are fixed throughout (${\bm \lambda}_k = [\lambda_1, \dots, \lambda_1]$) we have
\begin{align}\label{eq:accumulated_FI_non_adaptive}
	\mathcal{F}\measidx{k}^\text{acc}\left(\theta\right)&\overset{\text{apo}}{=}\mathcal{F}^\text{ind}(\theta;{\lambda_1}) \sum_{i=1}^{k}\sum_{\bm n_{i-1}} p({\bm n}_{i-1}|\theta;{\bm\lambda}_{i-1}) \nonumber\\
	&=k\mathcal F^\text{ind}(\theta;\lambda_1).
\end{align}
However, in the adaptive setup, the choice of parameters $\lambda_{i}$ can generally depend on the previously observed outcomes ${\bm n}_{i-1}$ so the accumulated Fisher information is not simply additive and \autoref{eq:FI_acc_def} should be used without simplification. Nonetheless, one can always bound it from above, by assuming that the control parameters are chosen to maximize the individual Fisher information. More rigorously, denote $\accentset{\ast}{\lambda}(\theta) = \arg\max_{\lambda} {\cal F}^{\rm ind}(\theta;\lambda)$, then we have
\begin{align}\label{eq:linear_FI_approx}
	\mathcal{F}\measidx{k}^\text{acc}\left(\theta\right)\leq k {\cal F}^{\rm ind}(\theta;\accentset{\ast}{\lambda}(\theta)),
\end{align}
which holds for all strategies considered here, including the greedy adaptive scheme. In fact, one expects the greedy adaptive strategy to saturate this bound in the asymptotic regime ($k\gg 1$). This is because, first, for large $k$ the estimate converges to the temperature $\theta$ that generated the data, causing the adaptive control parameter $\lambda_i$ to converge as well. Second, the greedy protocol selects the control parameter that maximizes the single-step information gain, which in the large-$k$ limit coincides with the value $\accentset{\ast}{\lambda}(\theta)$ that maximizes the Fisher information, see Sec. \ref{subsec:equivalence_of_K_F_optimization} in the Supplemental Material and~\cite{GlobalThermometry}.

An important consequence of both bounds is that the error asymptotically scales with $k^{-1}$ for all three strategies considered here, as established by \autoref{eq:uno_additive_FI}, \autoref{eq:accumulated_FI_non_adaptive}, and \autoref{eq:linear_FI_approx}. Thus, any other scaling is a transient behavior.

\begin{figure}
	\centering
	\includegraphics[]{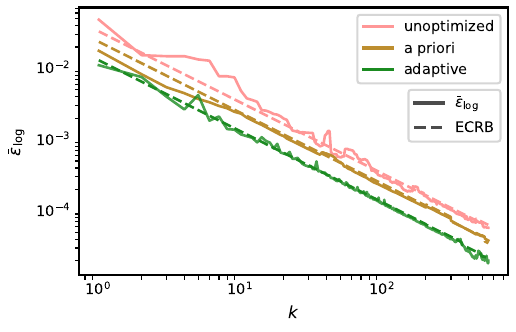}
	\caption{\textbf{ECRB and mean error trajectories.} The ECRB (dashed lines) is calculated according to the second line of \autoref{eq:bcrb_unbiased}, while $\mathcal{F}_k^\text{acc}(\theta)$ is calculated with \autoref{eq:uno_additive_FI} and \autoref{eq:accumulated_FI_non_adaptive} for the unoptimized and a~priori optimized protocols, respectively, and approximated by \autoref{eq:linear_FI_approx} for the adaptive protocol. Both quantities, the estimation error (solid lines) and the bound, are averages over the individual datasets corresponding to different temperatures and slightly different likelihood models. The averages are taken over eight (unoptimized), nine (a~priori), and nine (adaptive) datasets with different temperatures. On average, the adaptive protocol reduces the number of measurements required to reach an error of $\bar{\epsilon}_\text{log}<\num{1e-4}$ by a factor of approximately 2.5 compared with the unoptimized protocol.}
	\label{fig:BCRB_comparison}
\end{figure}

We now test the bound \autoref{eq:bcrb_unbiased} in our setup. In \autoref{fig:BCRB_comparison}, we show the mean logarithmic error averaged over multiple datasets at different temperatures (that is $\bar \epsilon _{\rm log}^k$, solid lines), with the different colors corresponding to the different protocols, and compare it with the expected Cramér--Rao bound (the right-hand side of \autoref{eq:bcrb_unbiased}) shown as dashed lines in the respective colors. For the unoptimized and a~priori optimized protocols, we use the additivity of the Fisher information to calculate the accumulated Fisher information according to  \autoref{eq:uno_additive_FI} and \autoref{eq:accumulated_FI_non_adaptive} respectively. For the adaptive protocol we use the upper bound on the Fisher information \autoref{eq:linear_FI_approx} to lower bound the right-hand side of \autoref{eq:bcrb_unbiased}. Since the likelihood model is subject to day-to-day fluctuations, the exact numerical value of the bound is specific to each dataset. We therefore also average the bound over all datasets. All protocols asymptotically saturate the expected Cramér--Rao bound~\autoref{eq:bcrb_unbiased}, including the adaptive protocol for which the upper bound \autoref{eq:linear_FI_approx} is used. Transiently, before the asymptotic regime is reached, the estimation errors can fall below the corresponding bound, as can be seen for the a~priori protocol (yellow), indicating that the assumptions underlying \autoref{eq:bcrb_unbiased} are not yet fully satisfied. In particular, this is consistent with a non-negligible estimator bias, whereas the bound assumes $b_k(\theta)=0$ for all $k$. For the unoptimized protocol, the estimation error initially lies above the expected Cramér--Rao bound and approaches it from above as the asymptotic regime is reached. The errors approach their respective bounds after different numbers of measurements for the three protocols. The onset of the asymptotic regime follows the same hierarchy observed for the other performance metrics: the adaptive protocol reaches it essentially from the beginning, followed by the a~priori optimized protocol, while the unoptimized protocol requires the largest number of measurements. Such behavior suggests that optimization reduces the finite-measurement effects associated with estimator bias more rapidly. This hierarchy is also reflected quantitatively in the convergence speed, with the adaptive protocol requiring, on average, approximately 2.5 times fewer measurements than the unoptimized protocol to reach the chosen error threshold of $\bar{\epsilon}_\text{log}<\num{1e-4}$.
As already stated before, the mean logarithmic errors exhibit an approximately inverse dependence on the number of measurements for all strategies.
\subsubsection{The Bernstein--von Mises theorem}
Furthermore, thanks to the Bernstein--von Mises theorem, the Fisher information also appears in describing the posterior distribution. Particularly, the theory predicts that the posterior will eventually be described by a normal distribution $\mathcal{N}$, centered at the estimate with a variance inversely proportional to the accumulated Fisher information~\cite{BvM_asymptotic_methods, 1998}. Formally, by the Bernstein--von Mises theorem, we have
\begin{gather}
	p(\theta|\boldsymbol{n}_k; \boldsymbol{\lambda}_k)\overset{k\gg 1}{\longrightarrow}\mathcal{N}[\eta_k,\sigma_k^2],\nonumber \\
	{\text{with }} \eta_k = \tilde{\vartheta}({\bm n}_k;\bm\lambda_k), \text{ and } \sigma_k^2 = {\mathcal{F}^{\rm acc}_k}(\tilde{\vartheta}({\bm n}_k;\bm\lambda_k))^{-1}.
	\label{eq:BvM}
\end{gather}
For $k \gg 1$, the estimator $\tilde{\vartheta}$ converges to $T$, and by continuity
of the Fisher information
$\mathcal{F}_k^{\mathrm{acc}}\big(\tilde{\vartheta}(\boldsymbol{n}_k;\boldsymbol{\lambda}_k)\big)
/\mathcal{F}_k^{\mathrm{acc}}(T) = 1 + \mathcal{O}(1/\sqrt{k})$.
To leading order, the posterior variance is therefore expected to satisfy
$\sigma_k^2 \simeq \mathcal{F}_k^{\mathrm{acc}}(T)^{-1}$,
independent of the observed outcomes~\cite{vanderVaart1998}.

\begin{figure}[t]
	\includegraphics[]{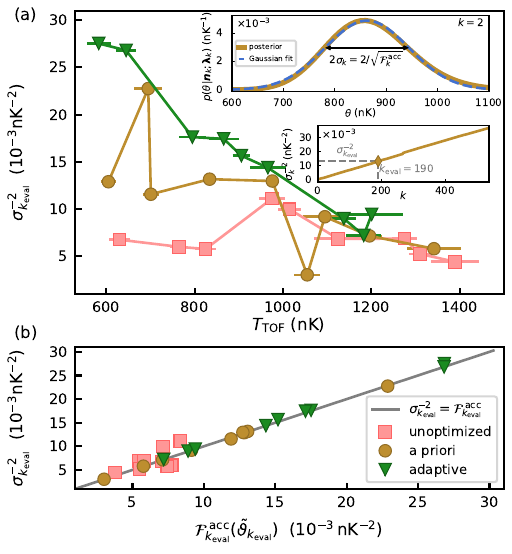}
	\caption{\textbf{Bernstein--von Mises theorem and Fisher information.} (a) The inverse posterior variance $\sigma_{k_\mathrm{eval}}^{-2}$ for different strategies at different temperatures (here $k_\mathrm{eval}=190$). It is obtained via Gaussian fits to the posterior PDFs. The upper inset shows a posterior and its Gaussian fit (dashed blue) for an exemplary dataset after $k=2$ in the a~priori optimized protocol (yellow). The lower inset shows the linear increase of  the inverse posterior variance $\sigma_k^{-2}$ for the same scenario. The diamond marker indicates the dataset shown in the insets. Connecting lines serve as guides to the eye. (b) Testing the Bernstein-von Mises theorem, \autoref{eq:BvM}. We plot $\sigma_{k_\mathrm{eval}}^{-2}$ versus the accumulated Fisher information ${\mathcal{F}}^{\rm acc}_{k_\mathrm{eval}}(\tilde{\vartheta}_{k_\mathrm{eval}})$ for all strategies and with various temperatures. The gray line shows the identity. In all three strategies, we observe that the posterior width follows tightly the predictions of the theory. Again, we are setting $k_\mathrm{eval}=190$.}
	\label{fig:FI_vs_T_and_model_comparison}
\end{figure}

In \autoref{fig:FI_vs_T_and_model_comparison}(a), we show the inverse posterior variance $\sigma_{k_\mathrm{eval}}^{-2}$ after $k_\mathrm{eval}=190$ measurements, for several runs of the experiment at different temperatures, for all three strategies. Analogous to the logarithmic error, the posterior variance establishes a hierarchy between the three different strategies, with adaptive performing the best, followed by the a~priori optimized and the unoptimized strategies. Note that the data points are for individual experimental datasets at fixed temperatures. It is thus natural that the hierarchy is not respected at some data points.
The protocols also differ in their temperature dependence. For the optimized protocols, $\sigma_{k_\mathrm{eval}}^{-2}$ decreases with $T$. This originates from the decreasing temperature sensitivity of the scattering rates $\Gamma_{m_F}$, see the inset of \autoref{fig:SE_comparison}(b) and Sec. \ref{subsec:optimization_equivalence} in the Supplemental Material. For the unoptimized protocol, $\sigma_{k_\mathrm{eval}}^{-2}$ remains approximately constant. This is consistent with the weaker temperature dependence of the Fisher information averaged over all $m=140$ unoptimized parameter tuples, shown by the red dotted line in \autoref{fig:optimization_equivalence}(b) of the Supplemental Material. The increased model sensitivity at low temperatures is partially compensated by the sampling of interaction times, which favors the slower dynamics at higher temperatures.

In the insets of \autoref{fig:FI_vs_T_and_model_comparison}(a) we examine the Bernstein--von Mises theorem for the a~priori optimized protocol in a sample run of the experiment ($T_\mathrm{TOF}=\SI{976\pm10}{\nano\kelvin}$, indicated by the diamond marker). In the upper inset, we extract the standard deviation of the posterior by fitting a Gaussian PDF and observe a nearly Gaussian shape already for $k=2$. 
The lower inset further shows that the inverse posterior variance increases linearly with $k$, that is $\sigma_k^{-2} \propto k$.
Lastly, in \autoref{fig:FI_vs_T_and_model_comparison}(b) we examine the Bernstein--von Mises theorem more broadly for all three strategies at various temperatures. Our results suggest that the posterior is well described by the theorem for all strategies, that is  $\sigma_k^{-2} \simeq \mathcal{F}_k^{\mathrm{acc}}(T)$. 

Asymptotic behavior is not guaranteed for numbers of measurements as small as those in our experiment. We attribute its early onset to the $\mu \approx 40$ probe atoms measured simultaneously in each round. By \autoref{eq:FI_ind_main}, they increase the Fisher information per measurement by a factor $\mu$. This also explains why the posterior is already close to Gaussian after $k = 2$ measurements. Other factors, such as the prior width, may contribute as well.

The Bernstein--von Mises theorem requires the likelihood to identify the temperature uniquely. This condition fails at parameters where population trajectories intersect, see Sec.~\ref{subsubsec:info_gain}. There, the Fisher information remains finite, since it only quantifies the local sensitivity of the likelihood to infinitesimal changes in temperature. Globally, however, two distinct temperatures yield the same likelihood. Repeated measurements at such parameters therefore lead to a bimodal posterior instead of a Gaussian one, see \autoref{fig:splitting_posterior} in the Supplemental Material. In this case, the information gain $\mathcal{K}$ can nearly vanish, even though the measurements still carry information. The reason is that $\mathcal{K}$, defined in \autoref{eq:K}, is an average over all possible outcomes $n_k$. For each outcome, it measures the squared logarithmic change of the estimate $\tilde\vartheta$, i.e., the change of the posterior mean of $\log\theta$. $\mathcal{K}$ thus vanishes if no outcome changes this mean, even though the posterior itself changes. A non-injective likelihood allows this. The update shifts probability towards both compatible temperatures, such that their effects on the mean of $\log\theta$ cancel.

	\section{Conclusion}
\label{sec:conclusion}

We have demonstrated Bayesian thermometry of an ultracold Rb bath using Cs atoms as non-equilibrium spin probes. We took the total number of measurements as the experimental resource. Within this framework, the expected error after each measurement has two contributions. The first is the error before the measurement, set by the current prior. The second, the \textit{information gain}, depends on the (potentially tunable) measurement parameters. Our first central result is a microscopic explanation of this dependence. We identified intersections between spin-population trajectories for different temperatures as the origin of poorly informative parameter regions. Based on this understanding, we designed and compared three protocols: an unoptimized, an a~priori optimized, and a greedy adaptive one. The two optimized protocols use the information gain to choose the control parameters. These are (i) the measured spin level and (ii) the interaction time. All three protocols agree with time-of-flight reference thermometry, up to a small systematic bias. However, they differ markedly in how efficiently they converge. The greedy adaptive protocol outperforms the a~priori optimized one, which in turn outperforms the unoptimized protocol. 

In the limit of many measurements, the estimation error of all three protocols scales as $1/k$ with the number of measurements $k$ and saturates the expected Cram\'er-Rao bound. Moreover, the posterior variance follows the inverse accumulated Fisher information, as predicted by the Bernstein--von Mises theorem. We observe this asymptotic behavior already within the moderate number of measurements of our datasets, which we attribute to the $\mu \approx 40$ probe atoms interrogated in each measurement. Since these atoms are uncorrelated, the Fisher information scales linearly with their number, i.e., our probe operates at the standard quantum limit (SQL). An exception occurs for measurements at parameters where population trajectories intersect: there, the likelihood is ambiguous between distinct temperatures and the posterior can become bimodal, such that the Bernstein--von Mises description no longer applies.

Working within a Bayesian framework allowed us to update the temperature estimate after every measurement, to monitor convergence shot by shot through the posterior, and to select the interaction time adaptively. Experimental fluctuations can likewise be incorporated into the likelihood as nuisance parameters and marginalized out. Here, a minimal likelihood model accounting for fluctuations of the total probe atom number already successfully described the measurement statistics, and it can be extended to additional experimental fluctuations. More generally, optimized single-atom quantum probes provide a minimally destructive and information-efficient route to thermometry in the ultracold regime.

Our results suggest several directions for future work. First, resources other than the number of measurements may matter in practice, such as the total laboratory time. In our setup, the a~priori optimized protocol is favorable in this respect, as it repeats an identical experimental sequence. In contrast, the adaptive and unoptimized protocols require a new control sequence after each measurement. The adaptive protocol additionally requires an optimization step. Such overheads are setup-specific. In practice, the optimal protocol therefore depends on the dominant experimental constraint. Second, the global character of the estimation problem may become more important for broader temperature intervals. The same holds for the choice of prior. A single a~priori optimized setting would then suit a smaller fraction of the interval. We would therefore expect a more pronounced separation between the strategies, favoring the adaptive protocol. Furthermore, the bath atom number currently enters the likelihood model as an independently measured parameter. A multi-parameter estimation framework~\cite{multi_parameter_estimation_review} could instead infer both the bath temperature and the bath atom number from the same probe data. This would also make it possible to study potential incompatibilities between the two parameters, since increasing the bath atom number and decreasing the temperature can lead to similar changes in the spin dynamics. Finally, the sensitivity of the probe could be enhanced by engineering the parameter-dependent dynamics, for example through Floquet driving of the trapping potential or the magnetic field~\cite{floquet_enhanced_sensing_luis,floquet_enhanced_sensing_maccone}.

	\begin{acknowledgments}
	\label{sec:template}
	We thank Luis A. Correa, Edward Gandar, Olga Movilla Miangolarra, Nessa Baum and Jonas Gutsche for providing insightful comments and carefully reading our manuscript. This work was supported by the Deutsche Forschungsgemeinschaft (DFG, German Research Foundation) via the Collaborative Research Center SFB/TR185 (Project-ID 277625399) and by the European Research Council (ERC) under the European Union’s Horizon Europe research and innovation programme (ERC Advanced Grant No. 101200776). S.B. acknowledges
	funding by Studienstiftung des deutschen Volkes. M.P.-L. acknowledges funding from  the ATRAE Program (Grant ATR2024-154621) funded by MICIU/AEI/10.13039/501100011033. This research was funded in part by the Austrian Science Fund (FWF) ["NOQUS", Grant DOI: 10.55776/PAT1969224].
\end{acknowledgments}

\section*{Author Contributions}
J.F. took and analyzed the experimental data. L.G. helped collect the experimental data. S.B. and S.H. helped run the experimental apparatus. A.G. performed the coupled-channel calculations, which provided the scattering cross sections. M.M. and M.P.-L. helped with the theoretical framework. A.W., M.M. and M.P.-L. conceived the research. J.F. wrote the initial manuscript draft. All authors contributed to the interpretation of the data, writing of the final manuscript, and critical feedback.

\section*{Data Availability}
The data that support the findings of this study are openly available at the following URL/DOI: https://doi.org/10.5281/zenodo.24390998
	
	\newpage
	\bibliography{./bibliography}{}
	
	\section{Supplemental Material}
\label{sec:supplemental}

\subsection{Experimental sequence}
\label{sec:experimental_sequence}

The experimental sequence begins with loading \Rb atoms from the background vapor into a 2D magneto-optical trap (MOT). A resonant pusher beam then creates a flux of transversely cooled atoms through a differential-pumping tube into an ultrahigh-vacuum (UHV) glass cell, where a 3D MOT captures and further cools the atoms. A compressed-MOT (cMOT) stage further compresses and cools the atomic ensemble. After the cMOT phase, the cold atomic cloud is transferred into a crossed optical dipole trap (ODT) and evaporatively cooled to ultracold temperatures. The evaporation is deliberately stopped before degeneracy is reached, and the final evaporation parameters are adjusted to prepare bath temperatures within the range investigated, namely $T=\SI{500}{\nano\kelvin}$--$\SI{1500}{\nano\kelvin}$. After evaporation to the desired final trap depth, the trap is recompressed to the same depth for all bath temperatures, ensuring identical trapping conditions across the different realizations. During evaporation, the atoms are prepared in the initial bath state $\ket{F=1,m_F=0}_\text{b}$ by optical pumping followed by a radio-frequency Landau-Zener transition. In parallel, \Cs atoms are captured from the background vapor and cooled in a MOT. We ensure low \Cs atom numbers by employing an umbrella MOT, which, after an initial loading phase, strongly increases the magnetic-field gradient to reduce the capture probability. The atomic density is further increased in a cMOT stage, followed by transfer into a second crossed ODT that shares the horizontal beam with the first ODT and is spatially separated from it by approximately \SI{200}{\micro\meter}. Here, we perform degenerate Raman sideband cooling (DRSC), which lowers the \Cs temperature to a few \si{\micro\kelvin} and pumps the \Cs atoms into their absolute internal ground state $\ket{F=3,m_F=3}_\text{p}$. Because of the finite efficiency of this pumping, we then remove all \Cs atoms that remain outside this ground state using $m_F$-selective microwave Landau--Zener transitions into the $F=4$ hyperfine manifold, followed by resonant laser push-out pulses. With this procedure, we prepare ensembles of approximately 40 probe atoms in their ground state with a fidelity of about 95\%. This degree of polarization is subject to day-to-day fluctuations. Therefore, the exact initial probe-state population vector $\bm q(\tau=0)$ is determined prior to the acquisition of each thermometry dataset and serves as the initial condition for solving the rate equation, ~\autoref{eq:rate}. The interaction between the two atomic species is initiated by species-selective transport of the \Cs atoms into the \Rb bath using a blue-detuned, one-dimensional optical lattice. The lattice is switched off during the interaction. To terminate the interaction and thereby set its duration $\tau$, the \Cs atoms are again confined axially by the lattice, and the \Rb bath is removed from the trap using a resonant laser push-out pulse. Subsequently, the \Cs atoms are detected by fluorescence imaging using a field-of-view extension (FOVE) scheme. To determine the internal state of the \Cs atoms, we employ the same state-selective microwave Landau--Zener transfer and removal procedure as for the initial-state cleaning. This removes all atoms except those in the selected state from the trap. Consequently, a single measurement does not determine the full vector of absolute spin-state populations, but only one of its components. The selected state $m_F$ therefore constitutes a control parameter of our thermometry protocols. For reference measurements of the \Rb cloud, we instead switch off the confining ODT, allow the cloud to fall and expand, and subsequently image it using absorption imaging. From the width and amplitude of a Gaussian distribution fitted to the absorption image, we infer the temperature estimate $T_\text{TOF}$ and the bath atom number $N_\text{b}$, respectively \cite{Schmidt2018Sep}.

\subsection{Maxwell-Boltzmann average $\langle \sigma v_\mathrm{c} \rangle_T$}
\label{subsec:MB}
To obtain the spin-exchange rate coefficients, we average the collision energy-dependent scattering cross sections over the relative velocities of thermalized probe and bath atoms. For both species at temperature $T$, the probability density of the relative speed $v_\mathrm{c}$ is
\begin{equation}
	p_\text{MB}(v_\text{c}, T)=4\pi\left(\frac{\mu_\text{red}}{2\pi k_\text{B}T}\right)^{3/2}v_\text{c}^2\exp\left(-\frac{\mu_\text{red} v_\text{c}^2}{2k_\text{B}T}\right),
\end{equation}
where $\mu_\text{red}=m_\mathrm{Cs}m_\mathrm{Rb}/(m_\mathrm{Cs}+m_\mathrm{Rb})$ is the reduced mass, $m_\mathrm{Cs}$ and $m_\mathrm{Rb}$ are the atomic masses, and $k_\mathrm{B}$ is Boltzmann’s constant. The thermally averaged rate coefficient for the transition $m_F\rightarrow m_F-1$ is
\begin{equation}
	\langle \sigma v_\text{c} \rangle_T=\int_0^\infty p_\text{MB}(v_\text{c}, T)\,\sigma(v_\text{c}, m_F)\,v_\text{c}\,dv_\text{c},
\end{equation}
with the corresponding spin-exchange cross section $\sigma(v_\mathrm{c},m_F)$. Multiplication by the spatially averaged bath density gives the collision rate $\Gamma_{m_F}(T)$ used in the rate equations.

\subsection{Time-of-flight thermometry}
\label{sec:TOF}
We independently determine the bath temperature and atom number by absorption imaging after releasing the \Rb cloud from the optical dipole trap. Following a time of flight ($t_\mathrm{TOF}=\SI{7}{\milli\second}$), the image is integrated along one spatial direction. The resulting line density is fitted with
\begin{equation}
	\rho_\mathrm{1D}(z)=A_\mathrm{Rb}\exp\left[-\frac{(z-z_0)^2}{2\sigma_\mathrm{Rb}^2}\right]+b,
\end{equation}
where z is the position along the remaining image coordinate, $z_0$ is the cloud center, $\sigma_\mathrm{Rb}$ is its Gaussian width, $A_\mathrm{Rb}$ is the amplitude, and b is a constant background. Assuming ballistic expansion and neglecting the initial cloud size, the temperature follows as
\begin{equation}
	T_\mathrm{TOF}=\frac{m_\mathrm{Rb}}{k_\mathrm{B}}\frac{\sigma_\mathrm{Rb}^2}{t_\mathrm{TOF}^2},
\end{equation}
with the \Rb atomic mass $m_\mathrm{Rb}$ and Boltzmann constant $k_\mathrm{B}$. Integrating the fitted Gaussian contribution yields the bath atom number,
\begin{equation}
	N_\mathrm{b}=\sqrt{2\pi}\sigma_\mathrm{Rb}A_\mathrm{Rb}.
\end{equation}
These measurements provide the reference temperature for the spin-probe estimates and independently fix the bath atom number entering the rate-equation model.

\begin{figure}
	\centering
	\includegraphics[width=\linewidth]{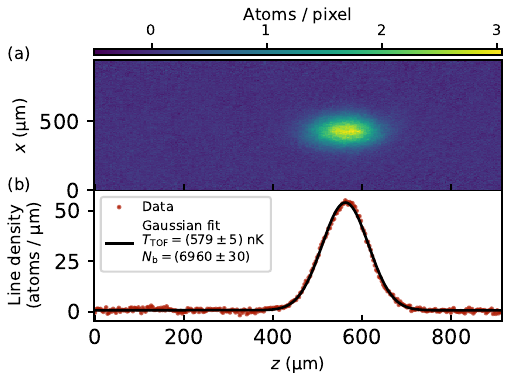}
	\caption{\textbf{Absorption imaging and TOF thermometry.} (a) Absorption image of the expanded Rb bath, with colors indicating the atom number per pixel. (b) Integrated line density (red points) and Gaussian fit (black line), yielding $T_\mathrm{TOF}=\SI{579\pm5}{\nano\kelvin}$ and $N_\mathrm{b}=\num{6960\pm30}$. The quoted uncertainties follow from the fit covariance.}
	\label{fig:placeholder}
\end{figure}

\subsection{Temperature estimation by $\chi^2$ fitting}
\label{subsec:chi_squared}

\begin{figure}[htb]
	\includegraphics[width=\linewidth]{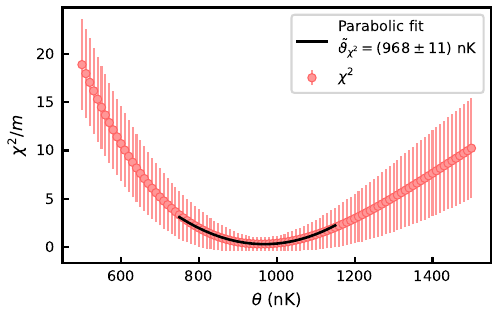}
	\caption{\textbf{Temperature estimation by $\chi^2$ fitting.} Normalized squared residuals $\chi^2/m$ versus bath temperature hypothesis $\theta$ for an unoptimized dataset, where $m=7N_\tau=140$ is the number of spin-population data points. Error bars indicate uncertainty estimates propagated from the measured relative atom numbers. The black line shows a parabolic fit near the minimum, yielding $\tilde{\vartheta}_{\chi^2}=\SI{968\pm11}{\nano\kelvin}$. The temperature uncertainty follows from the curvature of the $\chi^2$ curve using $\Delta\chi^2=1$.}
	\label{fig:chi_squared}
\end{figure}

For the unoptimized protocol, we additionally extract the bath temperature by fitting the measured averaged spin-population dynamics. After correction for probe-atom loss, the measured atom numbers are normalized and compared with rate-equation simulations. For a hypothetical temperature $\theta$, we calculate
\begin{equation}
	\chi^2(\theta)=\sum_{i=1}^{N_{\mathrm{\tau}}}\sum_{m_F=-3}^{3}\frac{\left[n_{\mathrm{rel}}(\tau_i,m_F) - q_{m_F}(\theta;\tau_i)\right]^2}{\left[\Delta n_{\mathrm{rel}}(\tau_i,m_F)\right]^2}.
	\label{eq:chi_squared_populations}
\end{equation}
Here, $N_{\mathrm{\tau}}$ is the number of sampled interaction times $\tau_i$, $n_{\mathrm{rel}}$ is the measured relative atom number, and $\Delta n_{\mathrm{rel}}$ is its propagated uncertainty. The modeled occupation probability is denoted by $q_{m_F}(\theta;\tau_i)$. The bath atom number, magnetic field, and initial spin distribution are fixed independently. An example of the resulting temperature dependence is shown in
\autoref{fig:chi_squared}. To extract the temperature estimate, we fit a parabola near the minimum,
\begin{equation}
	\chi^2(\theta) = a\left(\theta-\tilde{\vartheta}_{\chi^2}\right)^2+c,
	\label{eq:chi_squared_parabola}
\end{equation}
where $\tilde{\vartheta}_{\chi^2}$ is the fitted temperature estimate and $a$ is the quadratic coefficient. The uncertainty is given by $\Delta\tilde{\vartheta}_{\chi^2}= a^{-1/2}$ using $\Delta\chi^2=1$.

\subsection{Bimodal posterior}
\label{subsec:bimodal_posterior}

\begin{figure}
	\centering
	\subfloat{\includegraphics[width=\linewidth]{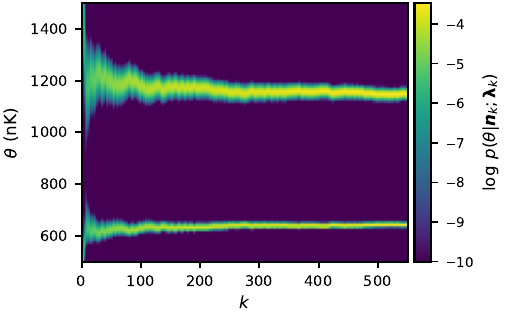}}
	\caption{\textbf{Posterior evolution of measurements at minimal information gain.}
		Every vertical slice corresponds to a posterior after measurement number $k$.
		Measurements performed at minimal information gain lead to a bimodal posterior PDF.}
	\label{fig:splitting_posterior}
\end{figure}

\begin{figure}
	\includegraphics[width=\linewidth]{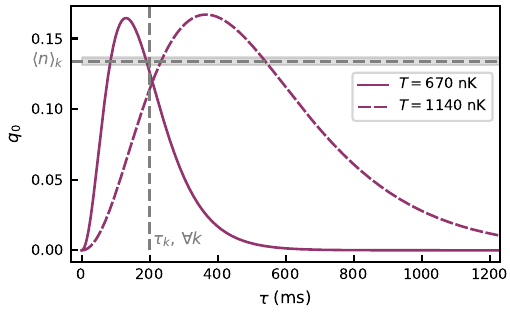}
	\caption{\textbf{Calculated population dynamics} $q_{0}$ for the measured state $\ket{m_F=0}$ at two different bath temperatures, $\theta_1\approx\SI{670}{\nano\kelvin}$ (solid) and $\theta_2\approx\SI{1140}{\nano\kelvin}$ (dashed). These temperatures correspond to the centers of the two contributions to the posterior shown in \autoref{fig:splitting_posterior}. The vertical dashed line marks the interaction time at which all measurements are performed, $\tau_k=\SI{200}{\milli\second}\ \forall k$, while the horizontal dashed line indicates the mean measured relative atom number $\langle n\rangle_k$ of these i.i.d. measurements. The gray shaded area indicates the standard deviation of these measurements. The population trajectories intersect at the interaction time used for the measurements and near the mean measured relative atom number.}
	\label{fig:crossing_pop_bimodal_prior}
\end{figure}

A natural question is how the Bayesian estimate evolves when measurements are performed at a parameter tuple $\lambda=(\tau, m_F)$ that yields minimal information gain. As an example, we choose measurements in state $\ket{0}$ after an interaction time of $\tau= \SI{200}{\milli\second}$, which corresponds to a minimum in \autoref{fig:K_and_two_mF}(a). To study that, we take measurements at these parameters and keep them fixed, similar to the a~priori optimized protocol, but at intentionally maximally uninformative parameters. The Bayesian evaluation is performed as before. The resulting posterior evolution is shown in \autoref{fig:splitting_posterior}.
The posterior develops a bimodal structure, with two pronounced peaks around $\theta_1\approx\SI{670}{\nano\kelvin}$ and $\theta_2\approx\SI{1140}{\nano\kelvin}$. These temperatures correspond to population trajectories that intersect at the chosen interaction time $\tau_k=\SI{200}{\milli\second}$ and near the mean measured relative atom number $\langle n\rangle_k$, see \autoref{fig:crossing_pop_bimodal_prior}. 
This shows that the bimodality of the posterior is caused by the ambiguities due to the intersections, and hence equal likelihood functions for different temperatures ${p(n|\theta;\lambda)=p(n|\theta';\lambda)}$. Since the lower lying peak of the posterior lies close to the reference temperature $T_\text{TOF}=\SI{661\pm33}{\nano\kelvin}$, a low mean information gain does not necessarily imply completely unproductive parameters. In cases when the two intersecting trajectories belong to very different temperatures, as shown here, a sufficiently narrow prior PDF can render the parameters still useful, as the measurement is globally ambiguous but can remain locally sensitive. A more informed prior can break this degeneracy.

\subsection{Convergence speed in units of laboratory time}
\label{subsec:k_t_in_lab_time}

\begin{figure}[htb]
	\includegraphics[width=\linewidth]{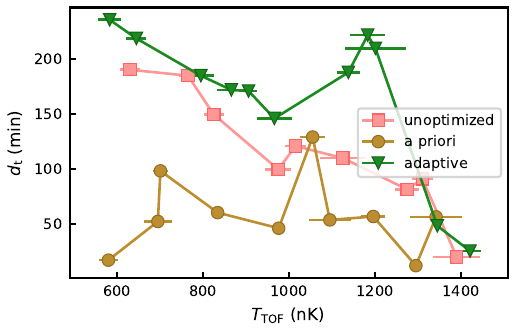}
	\caption{\textbf{Laboratory time required to reach the error threshold.} The laboratory time $d_\text{t}$ required for the estimation error to fall below the threshold $\epsilon_\text{t}$ marked in \autoref{fig:Comparison}(d), versus the reference temperature $T_\mathrm{TOF}$. The datasets are the same as in \autoref{fig:convergence_speed_vs_T}. Because adaptive optimization and the generation of a new experimental sequence add overhead, the adaptive protocol requires the most laboratory time in our setup.}
	\label{fig:conv_speed_lab_time}
\end{figure}

In the main text, we compare the convergence speed of the protocols by the threshold measurement number $k_t$, without accounting for the laboratory time required for each measurement. To include this experimental overhead, we determine the laboratory time $d_\text{t}$ required for the estimation error to fall below the same threshold $\epsilon_t=\num{1e-4}$. The resulting times are shown in \autoref{fig:conv_speed_lab_time} for the same datasets as in \autoref{fig:convergence_speed_vs_T}.

In terms of laboratory time, the a~priori optimized protocol performs best in our setup. Although the adaptive protocol requires the fewest measurements, the real-time optimization and generation of a new experimental sequence for changing measurement parameters increase its acquisition time. The unoptimized protocol is affected by similar sequence changes, but does not require the additional optimization step. Thus, the ranking based on laboratory time differs from that based on the number of measurements. This comparison is specific to the experimental overhead of our setup and the specific implementation of the optimization.

\subsection{Equivalence of information-gain and Fisher-information optimization}
\label{subsec:equivalence_of_K_F_optimization}
\begin{figure}[htb]
	\includegraphics[width=\linewidth]{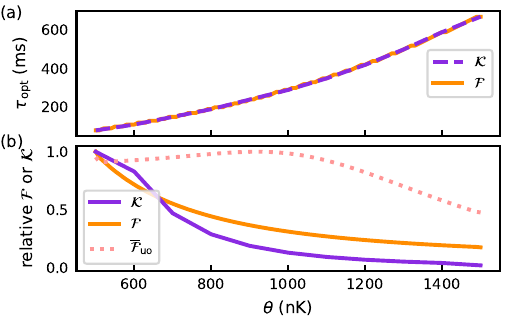}
	\caption{\textbf{Equivalence of information-gain and Fisher-information optimization.} (a) Optimal interaction time $\tau_\mathrm{opt}$ versus temperature $\theta$, obtained by maximizing the mean information gain $\mathcal K$ (purple dashed) or the individual-measurement Fisher information $\mathcal F^\mathrm{ind}$ (orange solid) for detection in $\ket{m_F=+3}$. The information-gain calculation uses narrow Gaussian priors centered at the respective temperatures. (b) Corresponding maximal values of $\mathcal K$ (purple) and $\mathcal F^\mathrm{ind}$ (orange). The red dotted line shows $\overline{\mathcal F}_\mathrm{uo}$, the Fisher information per measurement averaged over all $m=140$ parameter tuples of the unoptimized protocol: seven spin states and 20 interaction times between $\SI{0}{\milli\second}$ and $\SI{1200}{\milli\second}$. Each curve in (b) is normalized to its own maximum over the displayed temperature range.}
	\label{fig:optimization_equivalence}
\end{figure}

The mean information gain $\mathcal K$ accounts for the full prior distribution, whereas the Fisher information describes local temperature sensitivity. As the prior narrows, only the likelihood in a small temperature interval contributes significantly to the Bayesian update. Maximizing the expected information gain $\mathcal{K}$ therefore becomes equivalent to maximizing the local Fisher information. This equivalence is already expected from the global quantum
thermometry framework, which recovers local thermometry governed by the Fisher information in the limit of sufficiently localized posteriors~\cite{GlobalThermometry}. Figure~\ref{fig:optimization_equivalence}(a) demonstrates this agreement numerically for detection in $\ket{m_F=+3}$. Both criteria favor longer interaction times at higher temperatures, reflecting the slower spin-exchange dynamics. Their maximal values decrease with temperature but follow different dependencies, as shown in \autoref{fig:optimization_equivalence}(b). Thus, the agreement concerns the selected interaction times, while the two information measures retain distinct temperature dependencies.

For comparison, the red dotted line in \autoref{fig:optimization_equivalence}(b) shows the Fisher information averaged over all $m=140$ parameter tuples of the unoptimized protocol, denoted by $\overline{\mathcal F}_\mathrm{uo}$. These tuples comprise seven measured spin states and 20 interaction times between $\SI{0}{\milli\second}$ and $\SI{1200}{\milli\second}$. This average has a much weaker temperature
dependence than the Fisher information at the optimal interaction time. At low temperatures, the rapid spin dynamics leave many of the sampled interaction times in the nearly polarized, weakly informative regime. At higher temperatures, the slower dynamics spread the informative part of the evolution over more of the sampled times, partially compensating for the decreasing local temperature sensitivity. This averaging explains the qualitative difference between the protocols in \autoref{fig:FI_vs_T_and_model_comparison}(a), namely at a fixed measurement number, the inverse posterior variance follows the accumulated Fisher information and therefore exhibits a weaker temperature dependence for the unoptimized protocol.

\label{subsec:optimization_equivalence}

\subsection{The Fisher information for the measurement distribution}
\label{sec_FI_analatic}
Here, we derive the Fisher information of the atom-number likelihood used in \autoref{eq:FI_ind_main}. Starting from the statistical model introduced in Sec.~\ref{subsubsec:model}, we first show that the detected atom number follows a Poisson distribution $\mathcal{P}$ with mean $\mu q$ and then evaluate its Fisher information with respect to the temperature.

The probability distribution for $n$ is defined as
\begin{equation}
	p(n|q) = \sum_{N_0=n}^{\infty} \mathcal{P}_\mu(N_0) \mathcal{B}[n, N_0, q]
\end{equation}
where we used that $\mathcal{B}[n, N_0<n, q] = 0$. Here, the constituent distributions are defined as:
\begin{align}
	\mathcal{P}_\mu(N_0) &= \frac{\mu^{N_0}}{N_0!} \exp(-\mu), \\
	\mathcal{B}[n, N_0, q] &= \frac{N_0!}{n!(N_0-n)!} q^n (1-q)^{N_0-n}.
\end{align}
Substituting these into the summation yields:
\begin{equation}
	p(n|q) = \sum_{N_0=n}^{\infty} \frac{\mu^{N_0} e^{-\mu}}{N_0!} \left( \frac{N_0!}{n!(N_0-n)!} q^n (1-q)^{N_0-n} \right).
\end{equation}
Canceling $N_0!$ and pulling the terms independent of $N_0$ outside the sum:
\begin{equation}
	p(n|q) = \frac{e^{-\mu} q^n}{n!} \sum_{N_0=n}^{\infty} \frac{\mu^{N_0} (1-q)^{N_0-n}}{(N_0-n)!}.
\end{equation}
Let $k = N_0 - n$, then we have:
\begin{equation}
	p(n|q) = \frac{e^{-\mu} (\mu q)^n}{n!} \sum_{k=0}^{\infty} \frac{[\mu(1-q)]^k}{k!}.
\end{equation}
The summation is the Taylor series for $\exp[\mu(1-q)]$. Substituting this expression gives
\begin{equation}
	p(n|q) = \frac{e^{-\mu} (\mu q)^n}{n!} e^{\mu(1-q)} = \frac{e^{-\mu q} (\mu q)^n}{n!}.
\end{equation}
This demonstrates that $n$ follows a Poisson distribution with mean $\mu q$. 

To proceed further, we calculate the Fisher information for $q$. Then, we know that $F_\theta = [\partial_\theta q]^2 F_q$.
To this end, we calculate the log-likelihood:
\begin{equation}
	\log p(n|q) = -\mu q + n \log(\mu q) - \log(n!).
\end{equation}
Taking the first derivative of the log-likelihood with respect to $q$ we have
\begin{equation}
	\frac{\partial}{\partial q} \log p(n|q; \mu) = -\mu + \frac{n}{q}.
\end{equation}
Using the definition of the Fisher information as the expected value of the squared score, we obtain
\begin{equation}
	{\cal F}_q = \sum_{n=0}^{\infty} p(n|q)\left( \frac{\partial}{\partial q} \log p(n|q) \right)^2.
\end{equation}
By substituting the derivative into the expectation, we obtain
\begin{equation}
	{\cal F}_q = \sum_{n=0}^{\infty} p(n|q) \left( \frac{n - \mu q}{q} \right)^2  = \frac{1}{q^2} \sum_{n=0}^{\infty} p(n|q) (n - \mu q)^2.
\end{equation}
The summation is the variance of $n$. Since $n$ follows a Poisson distribution with mean $\mu q$, its variance is also equal to $\mu q$. Therefore:
\begin{equation}
	{\cal F}_q = \frac{1}{q^2} (\mu q) = \frac{\mu}{q}.
\end{equation}
From here, one can simply find that ${\cal F}_\theta = \frac{\mu (\partial_\theta q)^2}{q}$.

\end{document}